\documentclass{jfm}
\usepackage{graphicx}
\usepackage{epstopdf, epsfig}

\newcommand{\D}{\partial}
\newcommand{\la}{\langle}
\newcommand{\ra}{\rangle}

\newcommand{\bv}[1]{\boldsymbol{#1}}
\newcommand{\vn}{\bv{v}_n}
\newcommand{\vi}{\bv{v}_i}
\newcommand{\talpha}{\tilde{\alpha}}

\shorttitle{2D partially-ionized turbulence}
\shortauthor{S. J. Benavides and G. R. Flierl}

\title{Two-dimensional partially ionized magnetohydrodynamic turbulence}

\author{Santiago J. Benavides\aff{1}
  \corresp{\email{santib@mit.edu}} \and
  Glenn R. Flierl\aff{1}
	}

\affiliation{\aff{1}Department of Earth, Atmospheric, and Planetary Sciences, Massachusetts Institute of Technology, Cambridge, MA 02139, USA}

\begin{document}

\maketitle

\begin{abstract}
Ionization occurs in the upper atmospheres of hot Jupiters and in the interiors of gas giant planets, leading to magnetohydrodynamic (MHD) effects which couple the momentum and the magnetic field, thereby significantly altering the dynamics. In regions of moderate temperatures    the gas is only partially ionized, which also leads to interactions with neutral molecules. To explore the turbulent dynamics of these regions we utilize Partially-Ionized MHD (PIMHD), a two-fluid model -- one neutral and one ionized -- coupled by a collision term proportional to the difference in velocities. Motivated by planetary settings where rotation constrains the large-scale motions to be mostly two-dimensional, we perform a suite of simulations to examine the parameter space of 2D PIMHD turbulence and pay particular attention to collisions and their role in the dynamics, dissipation, and energy exchange between the two species. We arrive at, and numerically confirm, an expression for the energy loss due to collisions in both the weakly and strongly collisional limits, and show that, in the latter limit, the neutral fluid couples to the ions and behaves as an MHD fluid. Finally, we discuss some implications of our findings to current understanding of gas giant planet atmospheres.
\end{abstract}



\section{Introduction}\label{sec:intro}
The interior atmosphere of Jupiter-like gas giant planets is typically characterized by two dynamically distinct regions: a neutral outer envelope following the laws of hydrodynamics (HD), and a hot, ionized interior where the hydrogen transitions into a conducting, metallic liquid state which produces interactions between the momentum and magnetic field, following the laws of magnetohydrodynamics (MHD) \citep{Guillot2005,Liu2008,Stanley2010}. While the location of this transition might change depending on the planet's mass and age, the existence of the two regions is expected to be robust, given typical pressures, temperatures, and composition of gas giant planets. Attempts at modeling each region {\it individually} have successfully reproduced and helped with the understanding of a lot of the major observations of Jupiter and Saturn to this date. This includes, among many other things, the formation, characteristics, and dynamics of the jets and vortices \citep{Rhines1975,BUSSE1976255,Dowling1988,Dowling1989,Cho1996,Showman2007,Scott2007,Scott2008,Liu2008,Glatzmaier2009,LIAN2008597,LIAN2010373,Stanley2010,Schneider2009,Warneford2014,ONeill2015,Heimpel2016} as well as the generation and morphology of the magnetic field \citep{Stanley2010,Wicht2010,JONES2011120,Gastine2014,Jones2014,Rogers2017,DUARTE2018,DIETRICH201815}. However, in reality, the two regions are not completely independent of each other. The transition from neutral to fully ionized is believed to happen continuously as a function of the radius \citep{bagenal2006jupiter,Liu2008,Cao2017,Zaghoo2018}. This suggests a continuous transition from HD to MHD dynamics. Many of the modeling studies mentioned above have either ignored or parameterized the interaction of their modeling region with this transition region, where the dynamics begin to change. Some examples include the presence of a Rayleigh-like friction in General Circulation Models (GCMs) modeling the neutral region. This friction is coined ``MHD Drag'' and is supposed to account for the interaction between the jets and the metallic interior \citep{Liu2008,Glatzmaier2008,Schneider2009,Perna2010}. On the other hand, some modeling efforts of the interior MHD region have begun including the effects of variable conductivity as a function of radius, accounting for the steep drop off of conductivity as one approaches the neutral region \citep{JONES2011120,Gastine2014,Jones2014,DIETRICH201815}. Despite the relative success of these methods, some call into question these techniques \citep{Glatzmaier2008,Chai2016}, and many studies from both communities explicitly express interest in further understanding of the coupling between the HD and MHD regions \citep{Gastine2014,Jones2014,Heimpel2016,Chai2016}.

A better understanding of the transition region could also be important for understanding hot Jupiters, gas giant planets orbiting close to other stars. The outer atmospheres of hot Jupiters are expected to be ionized, due to both high temperatures and incident radiation from the nearby host star \citep{Batygin2010,Perna2010,Menou2012,Koskinen2014,Koll2017}, and a few studies have already implemented GCMs which include an electrically conducting atmosphere and a magnetic field \citep{Batygin2013,Rogers2014b,Rogers2017}. The resulting circulations depend significantly on the interaction of the flows with the magnetic field, which has implications for the interpretation of observed hot spots on hot Jupiters.

All previous numerical work described above has been carried out using either regular HD, which models the dynamics of an electrically-neutral fluid, or MHD, which models the dynamics of a \textit{fully-ionized}, electrically-conducting fluid, incorporating the interaction between the fluid and the magnetic field. These are called single-fluid or single-species models because they model only one type of molecule (either neutral or ionized). However, it is likely that this continuous transition occurs via {\it partial} ionization \citep{Zaghoo2018},
implying a coexistence of ionized and neutral molecules in this region, both following their own respective mean dynamics but occupying the same fluid volume and interacting via collisions.

The main goal of this work is to improve our understanding of the partially-ionized turbulent dynamics occurring in the transition region. A more rigorous understanding of the plasma physics and dynamical regimes there can shed light on the commonly used assumptions. To that effect, in section \ref{sec:pimhd} we introduce and explore Partially-Ionized MHD (PIMHD), a two-fluid model -- one neutral and one ionized -- coupled by a collision term proportional to the difference in velocities. Unlike the single-species (fully ionized or fully neutral) models, the coexistence of two species introduces a new frictional dissipation of energy and source of heating due to collisions between the differentially moving species.
In section \ref{sec:methods} we motivate and describe the approach to our study using numerical simulations of two-dimensional turbulence, whose results are presented and discussed in section \ref{sec:results}. Finally, in section \ref{sec:conclusions} we summarize the PIMHD numerical experiments and give tentative parameter values for Jupiter, making some connections to our motivating discussion in the current section.

\section{Partially-ionized magnetohydrodynamics}\label{sec:pimhd}

\subsection{PIMHD system}\label{subsec:pimhd}
In this study, we investigate \textit{incompressible} PIMHD with \textit{uniform species densities}, ignoring the complications of compression, stratification, and buoyancy. Incompressibility is expected to hold true in the deep atmospheres of gas giant planets,
although this is possibly less accurate for the outer regions of hot Jupiter atmospheres where partially-ionization is also relevant. Uniform species densities is harder to justify, and we admit that it would certainly play a role in real geophysical applications at large scales. Despite this, we make these assumptions because they simplify analysis and allow better comparison to previously established turbulence results. The PIMHD system then becomes the following:
\begin{subequations}
\begin{equation}
 \rho_n \left( \frac{\D}{\D t} + \vn \cdot \nabla\right) \vn = -\nabla p_n -\rho_i \rho_n \alpha (\vn-\vi) + \mu_n \nabla^2 \vn + \bv{F}_n, \label{eq:fulln} 
\end{equation}
\begin{equation}
 \rho_i \left( \frac{\D}{\D t} + \vi \cdot \nabla\right) \vi = -\nabla p_i + \rho_i \rho_n \alpha (\vn-\vi) + \bv{J} \times \bv{B} + \mu_i \nabla^2 \vi + \bv{F}_i, \label{eq:fulli}
\end{equation}
\begin{equation}
 \frac{\D \bv{B}}{\D t} = \nabla \times \left(\vi \times \bv{B}\right) + \eta \nabla^2 \bv{B} + \bv{F}_B, \label{eq:fullB}
\end{equation}
\begin{equation}
 \bv{J} = \frac{1}{\mu_0} \nabla \times \bv{B}, \quad \nabla \cdot \bv{B} = 0, \quad \nabla \cdot \vn = 0,\quad \nabla \cdot \vi = 0,\quad \rho_{tot} = \rho_i + \rho_n,\label{eq:div}
\end{equation}
\end{subequations}
where the subscripts represent an ionized (``\textit{i}'') or neutral (``\textit{n}'') component, potentially representing dissociated Hydrogen ions and recombined atoms, respectively \citep{Guillot2005,French2012}. For each species, $\bv{v}$ is the velocity, $\rho$ is the density, $p$ is the pressure, $\mu$ is the dynamic viscosity, $\bv{B}$ is the magnetic field, $\mu_0$ is the vacuum permeability, $\eta$ is the magnetic diffusivity, and $\bv{F}$ is a generic force which may include body or gravitational forces and other forms of dissipation, for example. These two species do not form two different layers -- they occupy the same space, i.e. at each point there are two velocities corresponding to $\vi$ and $\vn$.

The PIMHD system can be derived from the Boltzmann equations for singly charged ions, electrons, and a single neutral species \citep{Draine1986,Meier2011,Meier2012}. The derivation process is similar to that of the MHD system from an electron-ion plasma, the difference being the presence of extra collision and reaction terms between neutrals, ions, and electrons. One combines the momentum equations of each ionized species with Maxwell's Equations, ignoring any static charge sources (quasi-neutral approximation) and light waves (the electric field is set by Ohm's law). Further simplifications in the dynamics mainly come in ignoring the electron inertia and electron pressure, thus reducing the equations of motion to that of the magnetic field and of the center of mass between the ions and electrons. Making these approximations in MHD requires assuming that the electron to ion mass ratio is very small, as well as assuming that we're looking at length scales much larger than the ion or electron skin depths. We want to emphasize here that we are taking the MHD approximation for the ion species, which ignores many two-fluid effects commonly considered in astrophysical plasmas \citep{Ballester2018}. We believe that the MHD regime is valid in the system we are attempting to study, and we discuss the breakdown of some of these assumptions in subsection \ref{subsec:limits} and appendix \ref{app:x_lims}.
Partially-ionized models have been used in previous studies of the Earth's Thermosphere/Ionosphere, the Sun's Chromosphere  \citep{Khodachenko2004,Zaqarashvili2011,Khomenko2012,Leake2014,Song2017,Martinez-Sykora2017}, magnetic reconnection \citep{Lazarian2004,Smith2008,Malyshkin2011,Leake2012}, protoplanetary disks \citep{Balbus2009}, as well as molecular clouds and the interstellar medium \citep{Draine1980,Nakano1986,Falle2003,Oishi2006,Osullivan2007,Tilley2010,Meyer2014,Xu2016,Xu2017}. As far as we are aware, there have not been any studies looking at partially-ionized turbulence in the planetary atmosphere setting we are investigating here. Some of the main differences between the context of previous work and our deep atmosphere context is that the former typically deals with ionization fractions much smaller than one, as well as compressibility effects. We have also not found any study which does a systematic parameter space study of the turbulent PIMHD system which will be discussed in section \ref{sec:methods}.

The two fluids are coupled via collisions, represented by the second term on the right hand sides of equations (\ref{eq:fulln}) and (\ref{eq:fulli}). The coefficient $\alpha$ measures the strength of the coupling; it is approximately proportional to the collision cross-section of the two species and their thermal velocities, which in turn depends on the square-root of the temperature under equilibrium assumptions \citep{Draine1986,Meier2011,Leake2013}. For our purposes, $\alpha$ will be a parameter that we vary, although we will discuss possible realistic values of $\alpha$ in section \ref{sec:conclusions}. Looking at the energy equation (and ignoring other forms of dissipation) we see the effects of collisions on the energy of the individual species ($s \in \{i,n\}$) and as a whole:
\begin{subequations}
\begin{equation}
\frac{d E_s}{dt} = - \rho_i \rho_n \alpha \langle |\bv{v_s}|^2\rangle  + \rho_i \rho_n \alpha \langle \vn \cdot \vi \rangle, \label{eq:en_s}
\end{equation}
\begin{equation}
\frac{d E}{dt} = - \rho_i \rho_n \alpha \langle |\vi - \vn|^2\rangle, \label{eq:en_tot}
\end{equation}
\end{subequations}
where $\langle \cdot \rangle$ implies a domain integral, and $E_n = KE_n = \rho_n \langle |\vn|^2 \rangle /2$, $E_i = KE_i + E_B =  \rho_i \langle |\vi|^2 \rangle /2 + \langle |\bv{B}|^2 \rangle / (2 \mu_0)$, and $E = E_i + E_n$. Collisions conserve momentum, and the sign indefinite term in equation (\ref{eq:en_s}) tells us that the two species may exchange some energy via the collisions. Looking at equation (\ref{eq:en_tot}) we see that total energy is {\it lost} from these interactions, in a process we are calling ``collisional heating'' (CH) \citep{Vasyliunas2005}. In the absence of collisions and other dissipation terms the two species are uncoupled and behave as HD and MHD independently. Note that CH is something not accounted for in one-fluid models, and could therefore prove problematic if it is shown to be significant in these planetary systems.

Another new and important parameter of the PIMHD system is the ionization fraction, which we will denote by $\chi \equiv \rho_i/\rho_{tot}$. Since $\rho_{tot} = \rho_i + \rho_n$, we see that $(1-\chi) = \rho_n/\rho_{tot}$. The ionization fraction plays a role in the dynamics by influencing the acceleration and Reynolds number (which measures the relative strength of the advection term to the diffusion term) of each species, as well as the strength of the collision term. We note here that equations (\ref{eq:fulln})-(\ref{eq:div}) are not valid for ionization fractions which are strictly 0 or 1, as the fluid description breaks down as those limits are approached. In the very extreme limits, we expect certain assumptions made about time scale separations between molecular motion and mean motion are now longer valid as the densities become low and the mean free path for self-collisions becomes too large (e.g., leading to the breakdown of the thermal equilibrium assumption) \citep{Draine1986}. Furthermore, before this occurs, as one approaches $\chi \rightarrow 0$, Ohm's law must be altered, as will be discussed in subsection \ref{subsec:limits} and appendix \ref{app:x_lims}. In any case, since we expect transition regions in gas giant planets to span all values of ionization fraction from 0 to 1, we will not focus on extremes of ionization fraction in this work. Ionization fraction also modifies the magnetic diffusivity $\eta$. In MHD the origin of magnetic diffusivity comes from collisions between ions and electrons, however, in a partially ionized system we also have collisions between neutrals and electrons. Incorporating this in the expression for magnetic diffusivity gives us:
\begin{equation}\label{eq:eta}
\eta = \left(1+r\left(\frac{1-\chi}{\chi}\right)\right)\eta_{MHD}
\end{equation}
where $r$ is the ratio of cross sections of ion-electron collisions to neutral-electron collisions, and $\eta_{MHD}$ is the magnetic diffusivity for regular MHD. Typically we would expect $r \ll 1$ \citep{Leake2012}, implying a sudden increase in magnetic diffusivity for small values of $\chi$ (low ionization fraction). Since we won't be dealing with extreme values of $\chi$ in this work, this effect will not be important here.

\subsection{Limiting cases}\label{subsec:limits}
Before moving on to the numerical experiments in section \ref{sec:methods}, we find it useful to explore the limiting cases of the PIMHD system (while staying within the bounds of our assumptions which make the system valid, as discussed in the previous section). We will look at the extreme limits of $\alpha$ and comment briefly on the ionization fraction limits.

The relative strength of the collision and advection terms in equations (\ref{eq:fulln}) and (\ref{eq:fulli}) is determined by the ratio of the eddy turnover time to the time scale of collisions. We define the eddy turnover time in the usual way: $\tau_{eddy} \equiv L/U$, where $L$ is a typical length-scale of the system, and $U$ is a typical velocity. There are two time scales for collisions: $\tau_{coll,i} \equiv (\rho_n \alpha)^{-1}$ in the ion equation and $\tau_{coll,n} \equiv (\rho_i \alpha)^{-1}$ in the neutral equation. These are typically denoted in the literature as collision frequencies $\nu_{in} = \rho_n \alpha$ and $\nu_{ni} = \rho_i \alpha$. Since at the moment we are dealing with both $\chi \sim \mathcal{O}(1)$ and $(1-\chi) \sim \mathcal{O}(1)$, it is convenient to define $\tau_{coll} \equiv (\rho_{tot}\alpha)^{-1}$ so that $\tau_{coll,i} = \tau_{coll}/(1-\chi)$, and $\tau_{coll,n} = \tau_{coll}/\chi$. This allows us to define our second main nondimensional parameter (after $\chi$),
\begin{equation}
\tilde{\alpha} \equiv \frac{\tau_{eddy}}{\tau_{coll}} = \frac{L \rho_{tot} \alpha}{U}, \label{eq:talpha}
\end{equation}
which will determine the strength of the collision term compared to the advection term in the PIMHD system and will be a measure of how coupled the two fluids are. The limits in the cases below are really being applied to $\tilde{\alpha}$. Part of our goal is to predict the collisional heating, defined to be:
\begin{equation}
	CH \equiv \rho_i \rho_n \alpha \langle |\vi - \vn|^2\rangle. \label{eq:CH}
\end{equation}
The collisional heating is not only of interest for its implications to the astrophysical systems mentioned in section \ref{sec:intro}, but also because it is a measure of how \textit{coupled} the two fluids are. Eq. (\ref{eq:CH}) and its equivalent wavenumber spectrum (to be defined) will be of interest throughout the rest of this work. We expect two extreme regimes: (1) $\talpha \ll 1$, where the ions and neutrals are not coupled and therefore follow their own separate dynamics, colliding and exchanging energy as they do so. (2) $\talpha \gg 1$, where the ions and neutrals are extremely coupled meaning $\vi \approx \vn$, thereby lowering $CH$. We will look at the two regimes separately and discuss some findings in each.

Let's begin with the limit of $\tilde{\alpha} \ll 1$. In the case where $\tilde{\alpha} = 0$ exactly, we recover two uncoupled fluids behaving as HD and MHD with no collisional heating. But suppose now that $0 < \tilde{\alpha} \ll 1$. In the case of isotropic 3D turbulence, since both HD and MHD turbulence cascade energy to smaller scales \citep{Alexakis_Review}, we expect that the collisional heating will become negligible once $\talpha < 1$ and will eventually go to zero as we keep decreasing $\talpha$. The case of 2D PIMHD turbulence is quite different due to the presence of an inverse cascade of energy in the neutral species which causes energy to go to larger and larger scales. The energy builds until some dissipative force is able to balance it. For a finite-size domain of typical length $L_0$, as long as $\alpha > \mu_n/(\rho_i \rho_n L_0^2)$, this dissipative force is not the viscosity but the collisional heating. At steady state we expect collisions to balance the energy injected into the neutrals, which we call $I_n \equiv \la \vn \cdot F_n \ra$, and so $CH \approx I_n$ for $\talpha \ll 1$. This prediction becomes independent of $\alpha$ because, at steady state, all of the energy being injected at the forcing scale is expected to be dissipated away by collisional heating, and thus what changes for different values of $\alpha$ is not the dissipation rate, but the energy at the largest scales. Indeed, if we further assume that $|\vn|\gg|\vi|$, which should be the case for small values of $\talpha$ due to the inverse cascade of the neutrals but not the ions, then combining this result with equation (\ref{eq:CH}) and the definition of $E_n$ we can say further that
\begin{equation}
E \approx E_n \approx \frac{I_n}{2 \rho_i \alpha} \quad \textrm{for} \quad \talpha \ll 1. \label{eq:en_predic}
\end{equation}
The balance between $CH$ and $I_n$ has allowed us to approximately relate the energy of the neutral species with the energy injection rate, the ionization fraction, and the collision coefficient. If this limit of $\talpha$ were realized in the transition regions of gas giant planets, this could have possible implications for the saturation of the jets, whose formation are arguably attributed to the inverse cascade of kinetic energy in the presence of latitudinally varying rotation \citep{Rhines1975}. Their saturation speeds, and therefore the effective Rhines scale, could depend on the value of $\alpha$. We should note here that we have assumed that the Rossby deformation radius is much larger than the domain size, but we do not expect our results to change for finite Rossby deformation radius since the arguments above still hold. Namely, the energy will still be dissipated at large scales (although possibly not the largest available scales) where viscous dissipation is negligible. Apart from a possible large-scale friction, equation (\ref{eq:en_s}) tells us that collisions might be responsible for energy exchange between neutrals and ions. Indeed, when $|\vn|\gg |\vi|$, we might expect the second term on the right hand side to dominate the friction-like term for the ion kinetic energy equation, thus leading to an injection of kinetic energy from the neutrals into the ions. Given that the ions would then cascade energy to the small scales, this could prove to be another route for the energy to be taken away from large scales and dissipated efficiently.

In the high collisional limit, $\talpha \gg 1$, we are also able to make some predictions. The following results are valid in both two and three dimensions, as they don't depend on any turbulent cascade. It is possible to do an asymptotic expansion of our variables in $\talpha^{-1}$, since this will be very small. Doing so leads us to conclude that at $\mathcal{O}(\talpha)$, $\vi^{(0)} = \vn^{(0)} \equiv \bv{u}$. Thus, to lowest order, the two fluids are completely coupled and $CH = 0$. Going to $\mathcal{O}(1)$ gives us, for the momentum equation of each species,
\begin{subequations}
\begin{equation}
	\rho_{n} \left( \frac{\D}{\D t} + \vn^{(0)} \cdot \nabla\right) \vn^{(0)} = -\nabla p_n - \rho_i \rho_n \left(\vn^{(1)}-\vi^{(1)}\right) + \mu_n \nabla^2 \vn^{(0)} + \bv{F}_n, \label{eq:O1n} 
\end{equation}
\begin{equation}
	\rho_{i} \left( \frac{\D}{\D t} + \vi^{(0)} \cdot \nabla\right) \vi^{(0)} = -\nabla p_i + \rho_i \rho_n \left(\vn^{(1)}-\vi^{(1)}\right) + \bv{J} \times \bv{B} + \mu_i \nabla^2 \vi^{(0)} + \bv{F}_i, \label{eq:O1i} 
\end{equation}
\end{subequations}
The dynamics for $\bv{u}$ can be found by adding equations (\ref{eq:O1n}) and (\ref{eq:O1i}) and plugging in the fact that $\vi^{(0)} = \vn^{(0)} = \bv{u}$. If we instead divide each equation by their respective density and subtract one from the other, we get rid of the left hand sides and end up with an equation for $\vn^{(1)}-\vi^{(1)}$, which we can then use to get an expression for the next order correction of the collisional heating $CH$. We end up with the following equations, which are correct down to $\mathcal{O}(1)$ in $\talpha^{-1}$:
\begin{subequations}
\begin{equation}
\rho_{tot} \left( \frac{\D}{\D t} + \bv{u} \cdot \nabla\right) \bv{u} = -\nabla (p_n+p_i) + \bv{J} \times \bv{B} + (\mu_n+\mu_i) \nabla^2 \bv{u} + \bv{F}_n + \bv{F}_i, \label{eq:u} 
\end{equation}
\begin{equation}
\frac{\D \bv{B}}{\D t} = \nabla \times \left(\bv{u} \times \bv{B}\right) + \eta \nabla^2 \bv{B} + \bv{F}_B, \label{eq:b2}
\end{equation}
\begin{equation}\label{eq:ch_predic}
CH = \frac{\rho_n \rho_i}{\rho_{tot}^2} \frac{1}{\alpha} \left\langle \left|\frac{\bv{J}\times\bv{B}}{\rho_i} - \frac{\nabla p_i}{\rho_i} + \frac{\nabla p_n}{\rho_n} + \left(\frac{\mu_i}{\rho_i} - \frac{\mu_n}{\rho_n}\right)\nabla^2 \bv{u} + \frac{\bv{F}_i}{\rho_i} - \frac{\bv{F}_n}{\rho_n}\right|^2 \right\rangle.
\end{equation}
\end{subequations}
Note that $CH \propto \alpha^{-1}$, rather than the order one correction one might expect, because the cross terms in $|\vn-\vi|^2$ go away, leading to $|\vn-\vi|^2 = \alpha^{-2}|\vn^{(1)}-\vi^{(1)}|^2$, which one combines with $CH \propto \alpha|\vn-\vi|^2$ to give an $\alpha^{-1}$ dependence. 

Looking at the dynamical equation for $\bv{u}$ reveals that it behaves like an MHD fluid, but with total densities, pressures, viscosities, and forces. This suggests that, in the large $\talpha$ limit, the two fluids are coupled so that one-fluid models for the partially ionized region become valid. Since a turbulent fluid has a continuum of time-scales, it is more appropriate to think of $\talpha$ as the collisional strength at a typical scale $L$ (which has a corresponding typical eddy turnover time and velocity), implying that the highly coupled limit could potentially only be valid up to a certain scale, depending on the actual value of $\talpha$ and choice of $L$. We will investigate the scale-by-scale properties of PIMHD in section \ref{subsec:spatial}. Equation (\ref{eq:ch_predic}) gives us a prediction of the collisional heating based on order one quantities. It tells us that collisions are caused by an imbalance in acceleration between the two species. For example, the ions feel the Lorentz force while the neutrals don't; therefore the magnetic field accelerates the ions in a different direction than the neutrals, which would then cause collisions.
The results presented above, for both extremes of $\talpha$, will be tested and discussed in section \ref{sec:results}.

Now we will briefly mention the limiting cases in ionization fraction, $\chi$, while maintaining the assumption of large $\talpha$. Due to the separation of time-scales that occurs when $\chi \ll 1$ or $(1-\chi)\ll 1$, it is numerically challenging to simulate these parameter limits, and we did not explore these regimes in our work. We therefore leave the details of the full discussion to appendix \ref{app:x_lims} and summarize the results here. In the fully ionized limit, the ions don't feel the collisions and make up most of the fluid, thus making the dominant dynamics single-fluid MHD, albeit with a modified pressure and body force in the highly collisional limit due the fact that ions drag around neutrals. The low ionization limit is a bit more subtle. Certain assumptions in the derivation of MHD no longer hold, and thus the induction equation (\ref{eq:fullB}) must be modified to include the Hall term \citep{Pandey2008}, even at large scales. Furthermore, the neutrals, which dominate in this limit, still interact with the magnetic field indirectly via collisions with the ions, leading to what is called ``ambipolar MHD'' in the high collisional limit, wherein the collisional effects act to enhance the magnetic diffusion. The limit of both low ionization and large collisional coupling, where ambipolar MHD is valid, are particularly relevant for many astrophysical applications such as protoplanetary disks, molecular clouds, and the interstellar medium \citep{Balbus2009,Draine1980,Oishi2006,Tilley2010,Meyer2014,Xu2016,Xu2017}.

\section{Methodology}\label{sec:methods}
Based on this introduction and discussion of the PIMHD system, we will now describe the numerical experiments used to test some of the predictions from section \ref{sec:pimhd}, as well as study the fully turbulent system scale-by-scale.

Our numerical experiments will comprise solely of two-dimensional, incompressible PIMHD turbulence. 
We acknowledge that a series of rotating three-dimensional simulations would be ideal. However, a large parameter sweep consisting of around one hundred simulations at various ionization fractions, collision strengths, and Reynolds numbers would be computationally demanding. We choose instead to explore this parameter space for the two-dimensional case first, with the expectation that it will provide guidance for future three-dimensional studies. There are two main reasons why we believe our choice is not restrictive and does not make this study irrelevant to its more realistic counterpart. The first relies on the fact that the high-collisional results in section \ref{subsec:limits} are not dimension-dependent and thus should hold for both 2D and 3D turbulence. Secondly, those results which do depend on the dimensionality really only depend on the directions of the energy cascades, which we are respecting in our 2D simulations, since 3D rotating HD turbulence is expected to cascade energy to larger scales like 2D HD turbulence, and 3D MHD turbulence also cascades total energy to smaller scales.
Other simplifications will also be made for tractability of both the analysis and the numerics. 
In a realistic setting, assuming $\mu_s$ is not changing, the kinematic viscosity $\nu_s \equiv \mu_s/\rho_s$ will be a function of the ionization fraction and will thus affect the Reynolds number for each species. However, in this study we aim to isolate the effects of ionization fraction on the dynamics, and also wish to perform a large number of simulations. We therefore choose to keep the kinematic viscosity, and thus Reynolds number, constant (and equal) for each species.
For similar reasons we will fix $\eta$ so that the magnetic Prandtl number $Pr_m \equiv \nu_i/\eta = 1$ for all simulations. Although this is likely not true in realistic planetary settings (especially in the transition region where we expect the magnetic diffusivity to be large), we make this choice because the focus of this work is not to study the effects of $Pr_m$, which is purely an MHD parameter. This means we are ignoring the effects of the magnetic diffusivity's dependence on ionization fraction, seen in equation (\ref{eq:eta}).

Equations (\ref{eq:fulln})-(\ref{eq:div}) were solved in a doubly-periodic domain with side-length $2 \pi$ using a modification of a 2D MHD code written by Prof. Pablo Mininni at the University of Buenos Aires, Argentina. The code was extended to include a neutral species and thus solve the PIMHD equations. It is a standard parallel pseudo-spectral code with a fourth-order Runge-Kutta scheme for time integration and a two-thirds dealiasing rule. More details on the parallelization can be found in \cite{Gomez2005}. All runs started from random initial conditions, were continuously forced, and were carried out long enough so that a statistically steady state was reached. All data was averaged at this state unless otherwise stated. A $512^2$ resolution was used for most of the experiments to explore the parameter space, with some $1024^2$ runs to ensure that the results are not resolution-dependent and to explore higher Reynolds numbers.

In an attempt to approach more realistic forcing mechanisms in geophysical flows, where convection or baroclinic instability might convert potential energy to kinetic energy, the forcing of each species is proportional to its density: $\bv{F}_s = \rho_s \bv{f}$, where $s \in \{i,n\}$. The forcing function $\bv{f}$ is identical for both species. $\bv{f}$ is random, white-in-time, and spectrally focused around wavenumber magnitude $k_f$, an input parameter. More specifically, at each time step, a wavenumber $\bv{k}_r$ of magnitude $k_f$ is chosen at random, and $\bv{\hat{f}}(\bv{k})$ (Fourier transform of $\bv{f}$) is set to zero everywhere except for at $\bv{k}_r$ where it has the magnitude $f_k/\sqrt{\Delta t}$, $f_k$ being another input parameter. This has the effect of setting the energy injection rate of each species to be $I_i = \rho_i f_k^2$ for ions and $I_n = \rho_n f_k^2$ for neutrals; see \cite{Chan2012} for more details. Constant energy injection rate into the system is ideal for studying situations in which there is little to no large-scale dissipation and hence a large-scale condensate forms \citep{Gallet2013}. The magnetic field was also forced using the function $\bv{f}$, but with a different random seed, and the magnetic energy injection rate was set to be $I_B = I_i/4$ for all runs. For most runs we set $k_f = 8$, so that both inverse and forward cascades could be resolved. However, $k_f = 4$ and $k_f = 32$ runs were carried out as well, to better resolve the forward and inverse cascade, respectively. In all runs, given different values of $k_f$, $I_s$ was chosen so that $u_f \equiv |\vi(k_f)| \sim |\vn(k_f)| \sim 1$ for both species.

In an attempt to better resolve inertial ranges while forcing at intermediate and larger wavenumbers, hyperviscosity, $(-1)^{p+1} \nabla^{2p}$, was used in all runs and for all three fields, replacing the regular viscosity and regular magnetic diffusivity seen in equations (\ref{eq:fulln})-(\ref{eq:fullB}). As long as the value of $p$ is not very large, hyperviscosity has been shown to have no significant effect on the turbulent properties of 3D turbulence, and we expect the same to be the case for our work \citep{Agrawal2020}. The value of $p$ was set to 2 in all runs except for those where $k_f = 32$, in which case $p = 4$. Furthermore, due to the inverse cascade of the square of the magnetic vector potential  $|A|^2$ \citep{Alexakis_Review}, where $\nabla \times (A\bv{\hat{z}}) = \bv{B}$, we chose to include hypoviscosity in equation (\ref{eq:fullB}) by adding $\eta_-\nabla^{-2} \bv{B}$. This acts to dissipate magnetic energy only at the largest scales and thus avoids the slow-forming condensate of $|A|^2$, making our simulations reach steady state faster. A condensate of the magnetic vector potential could possibly affect the dynamics of the ion species, but we consider this to be a purely MHD effect and thus not a focus of our work. The coefficient $\eta_-$ was chosen to ensure that the magnetic energy at the largest scales ($|\bv{k}|=1$) was smaller than that at the next largest scale, thus avoiding the formation of a condensate. 

In our simulations, we divided the momentum equations by $\rho_{tot}$ and absorbed it into the definition of our variables. Thus, in our simulations, $\bv{B} = \bv{B}/\sqrt{\mu_0 \rho_{tot}}$ and $\alpha = \rho_{tot} \alpha$ (making it equivalent to collision frequency), the latter being another input parameter. Doing this lets us directly employ the ionization fraction $\chi$ in the numerical integration of the equations. This means that $I_n = (1-\chi)f_k^2$ and $I_i = \chi f_k^2$. Using the four-fifths law, the typical velocity based on input parameters is $u_f = (f_k^2/k_f)^{1/3}$, which was maintained at 1 for all runs. Therefore, the eddy turnover time $\tau_{eddy} = (k_f f_k)^{-2/3}$. Furthermore, we define the numerical version of $\talpha$ based on input parameters:
\begin{equation}\label{eq:talpha_num}
\talpha \equiv \frac{\alpha}{k_f u_f} = \frac{\alpha}{(k_f f_k)^{2/3}}.
\end{equation}
We can also define a parameter analogous to the Reynolds number in terms of our simulation input parameters:
\begin{equation}
	Re = \frac{f_k^2}{\nu k_f^{-2p+2/3}},
\end{equation}
where $p$ is the power of the hyperviscosity. Since we are keeping $\nu$ the same for both species, we do not distinguish between $Re_i$ or $Re_n$ and simply call it $Re$. Note that our use of hyperviscosity means that the parameter we call $Re$ is not exactly a Reynolds number. However, $Re$ can still be seen as the ratio of eddy to diffusive time-scales, and thus measures the relative importance of advection compared to diffusion. From now on, during any discussion of our numerical results, all references to these variables will be the numerical versions defined above. A summary of our runs can be seen in table \ref{tab:runs}.
\begin{table}
  \begin{center}
\def~{\hphantom{0}}
  \begin{tabular}{ccccccc}
   Set Name & $k_f$ & $\chi$         & $\talpha$               &$Re$     & $p$ & $N$ \\[3pt]
     	K4	&  4    & [0.5]       & [3.15e-3, 3.15 $-$ 6.3] & 124,015 & 2   & 1024 \\
     	K8	&  8    & [0.1-0.99]  & [1.6e-4 $-$ 1.6e3]      & 1,938   & 2   & 512 \\
     	K8R	&  8    & [0.5]       & [3.0e1 $-$ 1.6e2]       & 15,502  & 2   & 1024 \\
     	K32	&  32   & [0.5]       & [1.5e-2 $-$ 7.9e-1]     & 1,938   & 4   & 512 \\
  \end{tabular}
  \caption{A summary of the runs performed for this work. $N$ is the resolution of the simulation and $p$ is the hyperviscosity exponent.}
  \label{tab:runs}
  \end{center}
\end{table}
Since we are numerically integrating the two-fluid equations, when $\chi$ becomes very small we begin to encounter time-scale separation issues in the equations of motion, causing numerical difficulties \citep{Falle2003,Osullivan2007}. Furthermore, at extremely small values of $\chi$ we would be forced to alter the equations of motion to those seen in equations (\ref{eq:ambi}) and (\ref{eq:ambi2}). For these reasons, and the fact that we are interested in spanning ionization fractions from zero to one, we chose to ignore extreme values of ionization fraction in our work. Set K8 was the most extensive set, with runs spanning $\talpha =$ [1.6e-4 $-$ 1.6e3]. Since each set has a specific, fixed $k_f$ and $f_k$, $\talpha$ was modified by varying the numerical value of $\alpha$. For each value of $\talpha$ six runs were performed where we varied $\chi$ from 0.1 to 0.99. The other sets were all run at $\chi = 0.5$ and focused mainly on the effect of the collisional strength on the dynamics. 

In the forthcoming section, we will present and analyze the results of our simulations. We divide the analysis into two subsections: subsection \ref{subsec:global}, `global' analysis, where we investigate the behavior of volume-averaged statistics and compare to the predictions in subsection \ref{subsec:limits}.  Subsection \ref{subsec:spatial}, `spatial' analysis, investigates the scale-by-scale effects of the collision strength on the two fluids.

\section{Results}\label{sec:results}
\subsection{Global}\label{subsec:global}
\begin{figure}
	\centering{
		\includegraphics[width=0.32\textwidth]{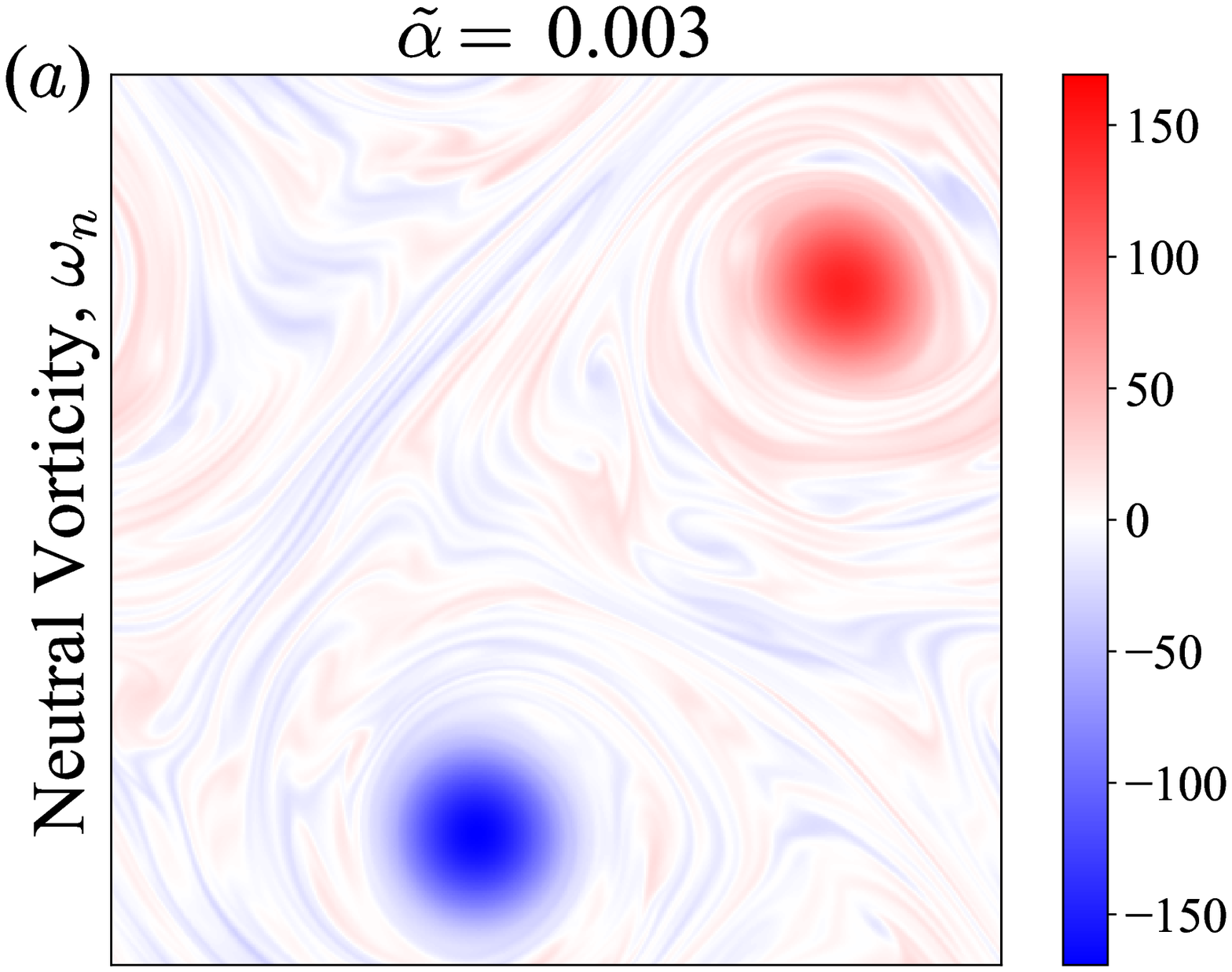}
		\includegraphics[width=0.32\textwidth]{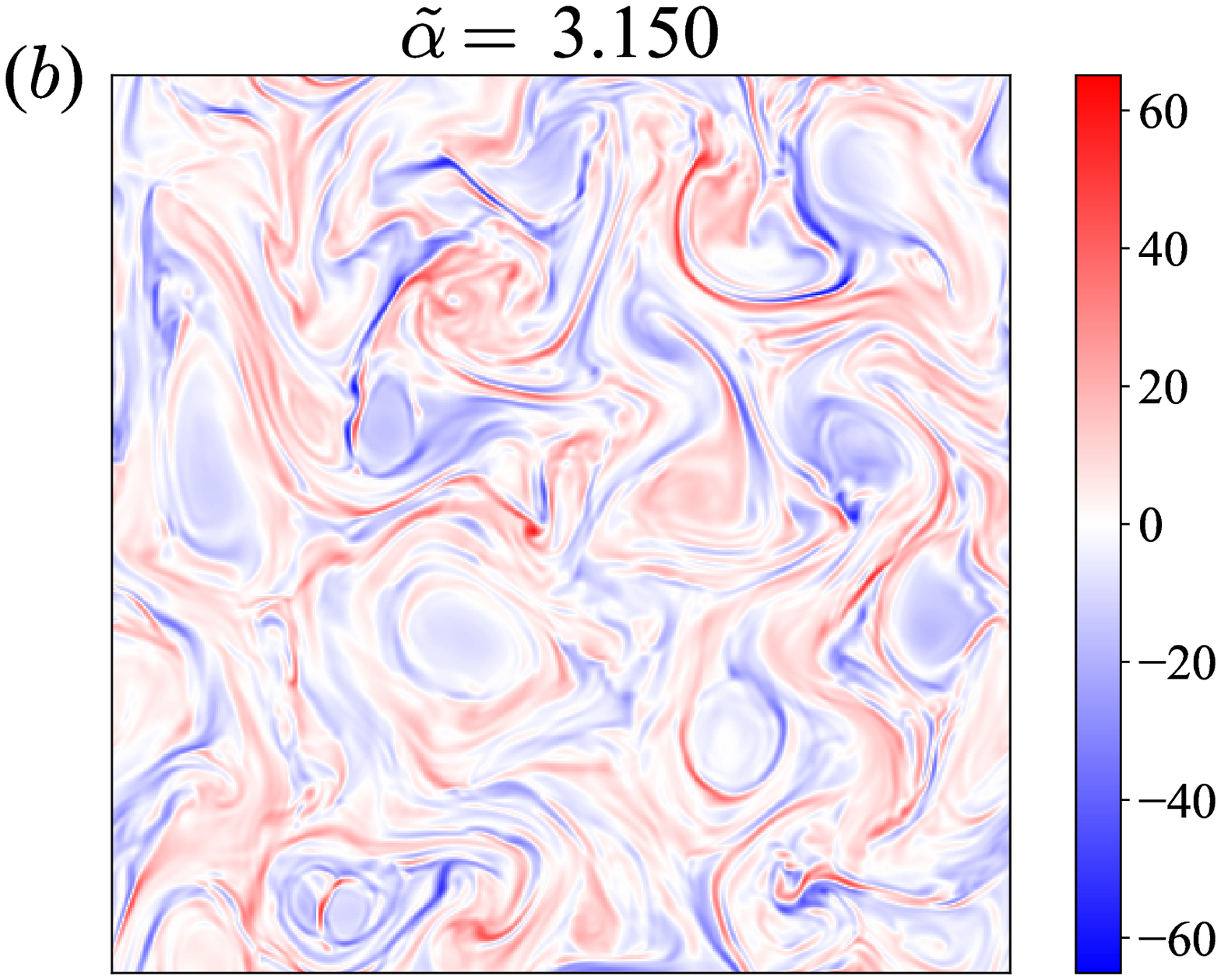}
		\includegraphics[width=0.32\textwidth]{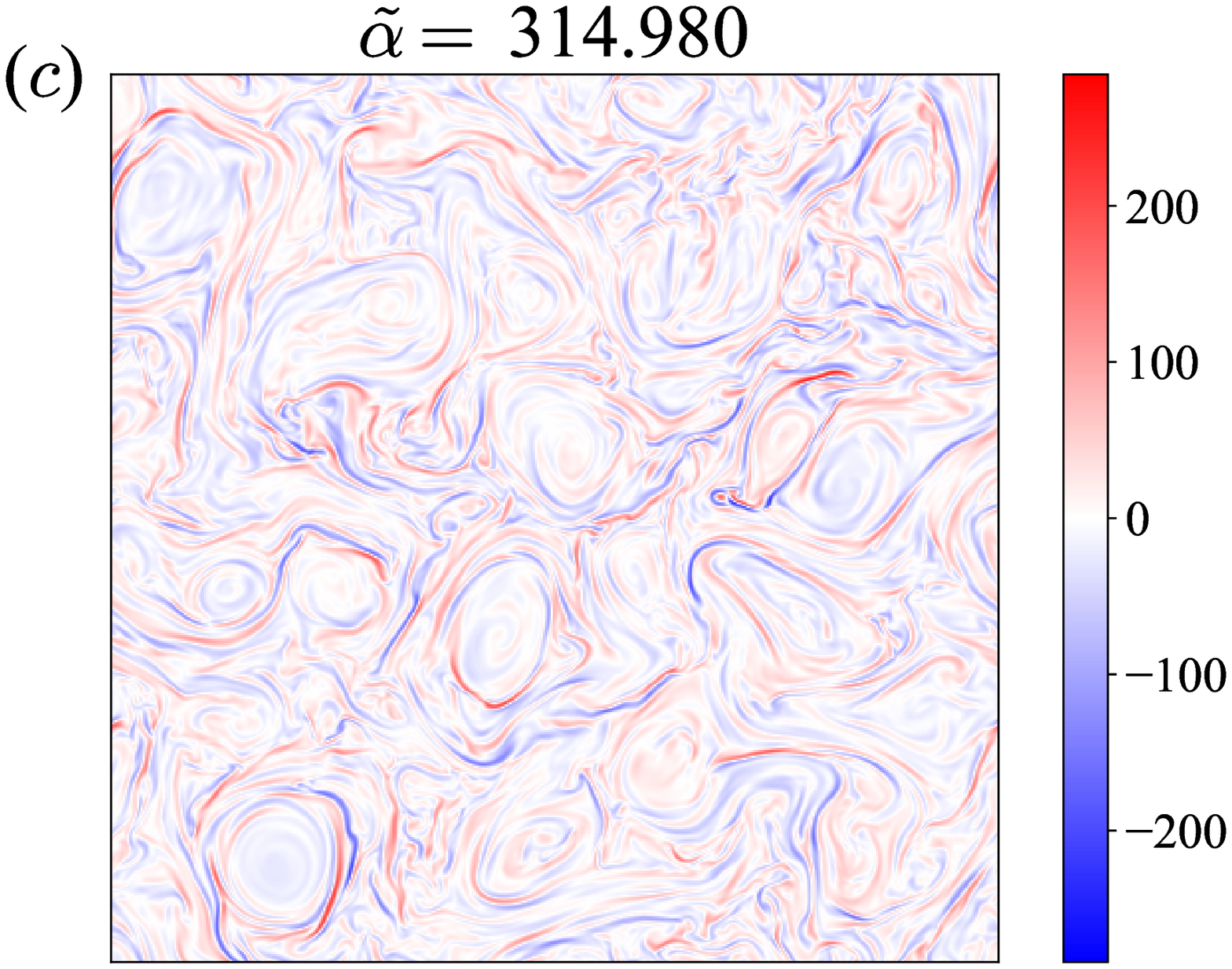}		\includegraphics[width=0.32\textwidth]{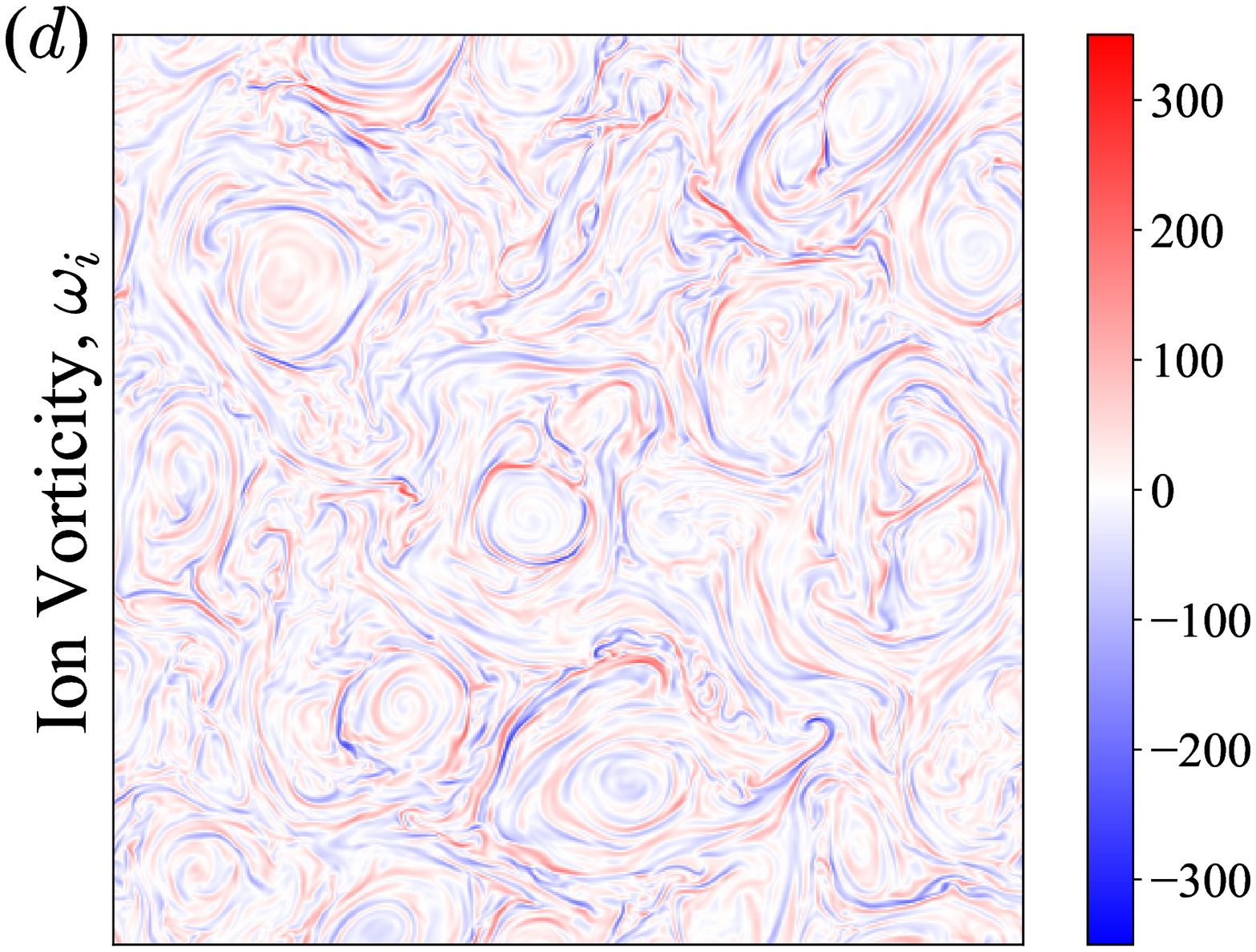}
		\includegraphics[width=0.32\textwidth]{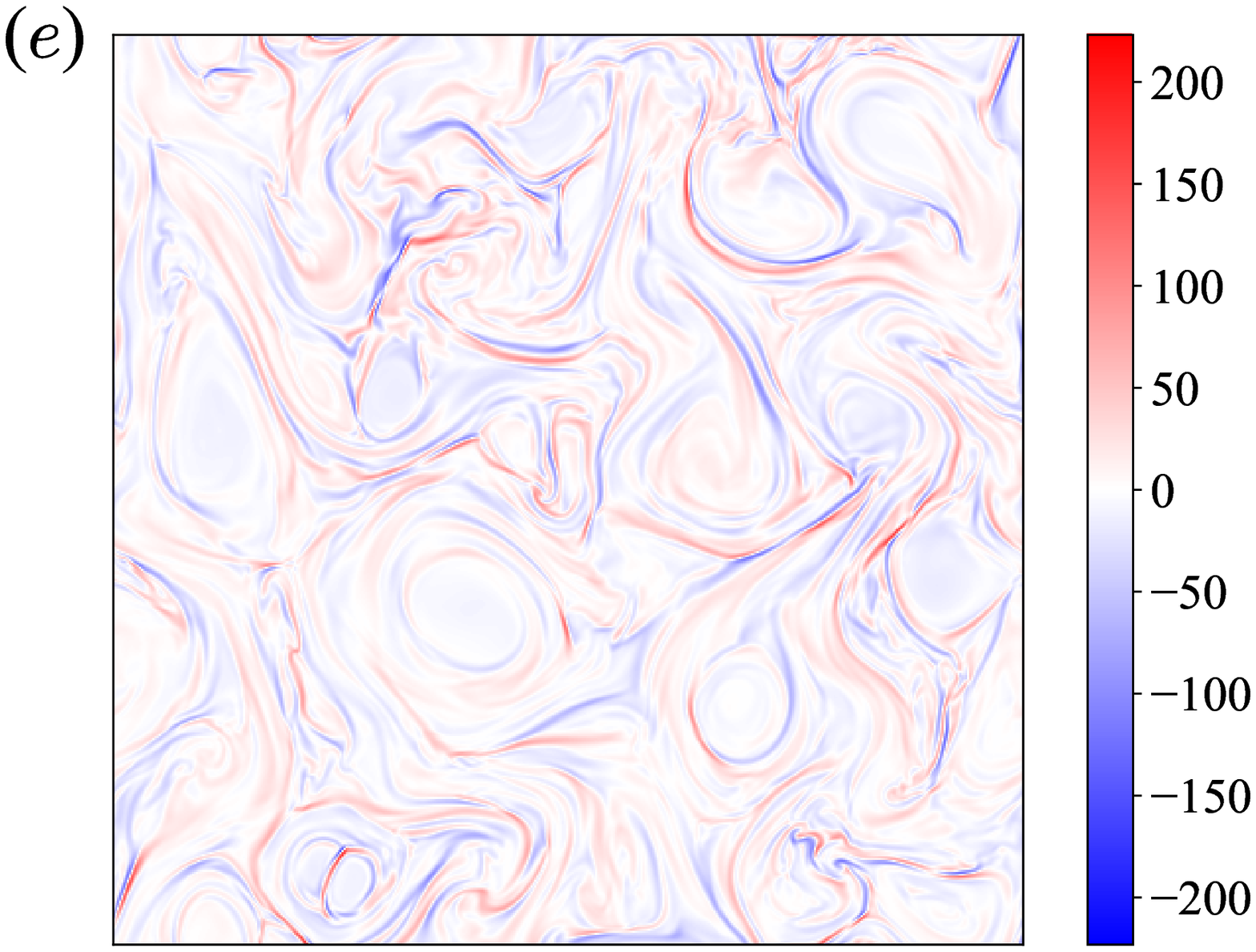}
		\includegraphics[width=0.32\textwidth]{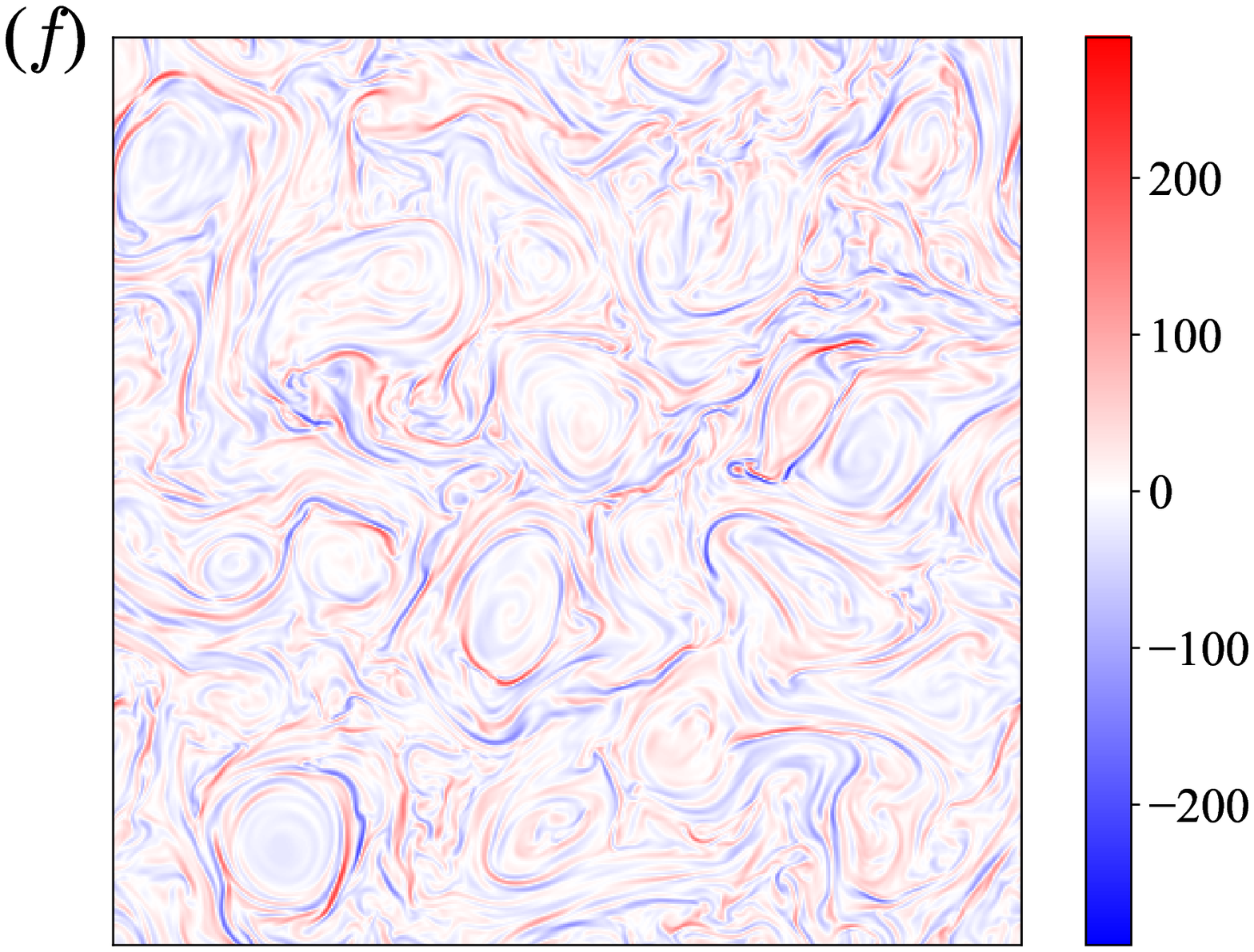}
	}
	\caption{
		(Colour online) Snapshots at steady state from K4 runs of the vorticity for each species, $\omega_s \equiv \nabla \times \bv{v}_s$, representing, from left to right, $\talpha \ll 1$, $\talpha \sim 1$, and $\talpha \gg 1$. Subfigures (\textit{a})-(\textit{c}) show the neutral vorticity for the three values of $\talpha$ whereas (\textit{d})-(\textit{f}) show the ion vorticity for those runs.}
	\label{fig:flow}
\end{figure}

In subsection \ref{subsec:limits} we saw that the collisional strength $\talpha$ sets the degree of coupling between the two fluids -- in the small $\talpha$ limit we expect the two fluids to move independently whereas, in the large $\talpha$ limit, they should be almost identical, following the MHD equations (\ref{eq:u}) and (\ref{eq:b2}). In figure \ref{fig:flow} we see three snapshots of the vorticity for each species, $\omega_s \equiv \nabla \times \bv{v}_s$, representing, from left to right, $\talpha \ll 1$, $\talpha \sim 1$, and $\talpha \gg 1$, all from the K4 runs. The top row shows the neutral vorticity for each run, whereas the bottom row shows the ion vorticity for those runs.
By visual inspection of the left-most column, subfigures \ref{fig:flow}(\textit{a}) and \ref{fig:flow}(\textit{d}), with $\talpha = 0.003$, one can indeed see that for small $\talpha$ the two species behave as they would in a completely uncoupled regime. The neutral vorticity shows clear signs of the 2D HD inverse cascade of kinetic energy given by the two large-scale vortices, whereas the ion vorticity shows many filamented structures, typical of 2D MHD turbulence.  In this regime, we can approach the question of whether or not a one-fluid description is adequate for the proper determination of the dynamics by going into the center-of-mass frame, typically done in other plasma settings (e.g. when deriving MHD itself). We define $\bv{V} \equiv \chi \vi + (1-\chi) \vn$ and $\bv{D} = \vi - \vn$. The question now becomes: is it possible to only account for the dynamics of $\bv{V}$ without knowing or integrating the dynamics of $\bv{D}$? Although it is not shown here, the simulations reveal that in the $\talpha \ll 1$ regime this is generally not possible -- that is, the dynamics of $\bv{V}$ are partially determined by $\bv{D}$ and vice versa. However, as $\chi \rightarrow 0$ or $\chi \rightarrow 1$, we find that a one fluid description ($\bv{V}$ only) is sufficient, as one might expect. This was shown by comparing the relative magnitudes of $\bv{V}$ and $\bv{D}$ and noting when $|\bv{D}|\ll|\bv{V}|$.

As we increase $\talpha$ so that it is order one, we see the neutral species begin to lose the large scale vortices, which are dissipated away by collisional heating. In the center column, subfigures \ref{fig:flow}(\textit{b}) and \ref{fig:flow}(\textit{e}), at $\talpha = 3.15$, we no longer see obvious 2D HD behavior from the neutral species, but it also does not appear to be of a similar nature to the ion vorticity. Later scale-by-scale analysis will reveal that this regime is where the highest collisional heating is found and that most of the energy injected into the neutrals will be transferred to the ions or simply dissipated away. On the right panels, subfigures \ref{fig:flow}(\textit{c}) and \ref{fig:flow}(\textit{f}), we see the case of $\talpha = 314.98$, and note immediately that the two fluids look identical from visual inspection. The two fluids are coupled and so the neutral species is behaving like an MHD fluid, as was predicted. 

\begin{figure}
	\centering{
		\includegraphics[width=0.49\textwidth]{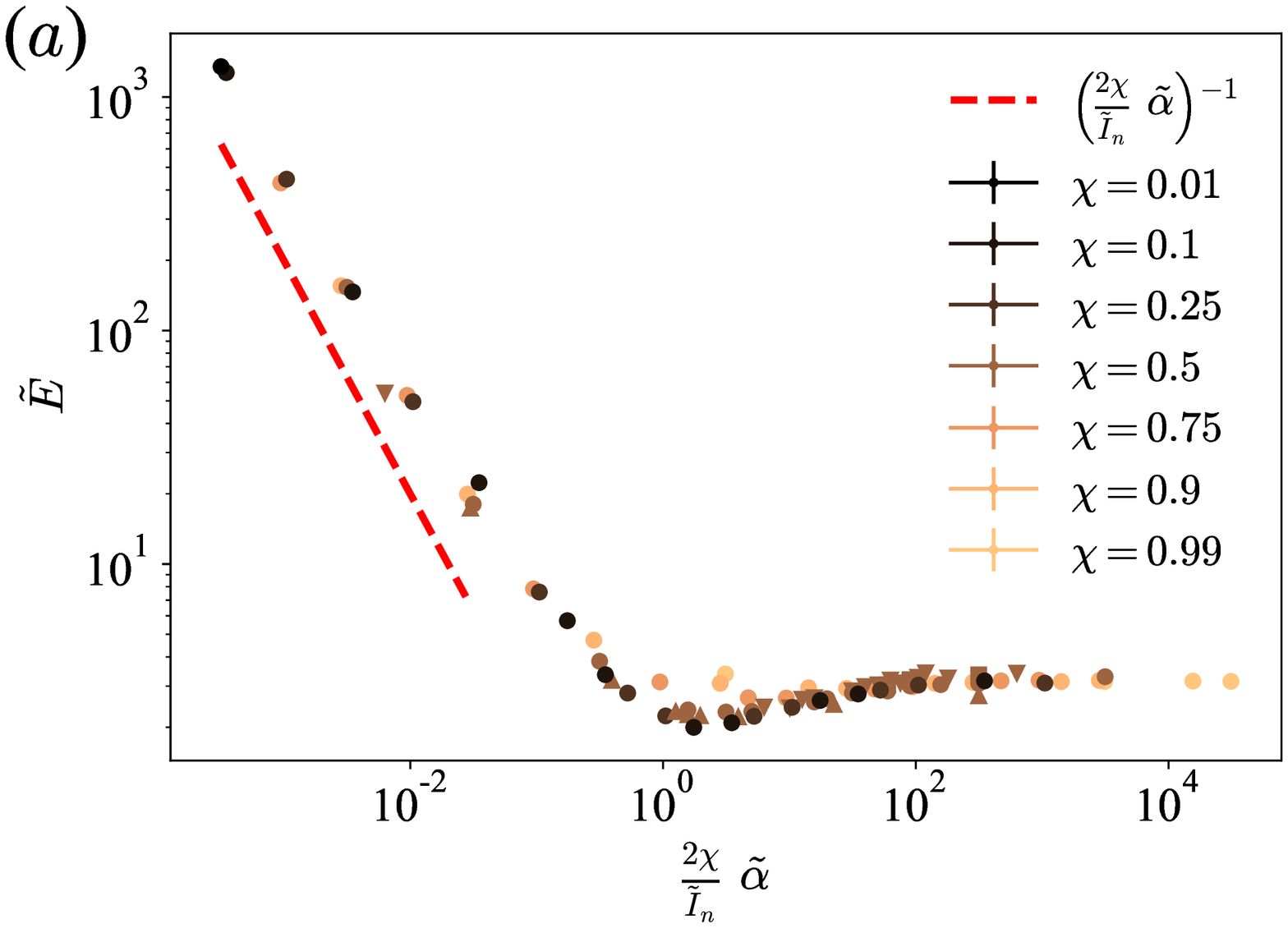}
		\includegraphics[width=0.49\textwidth]{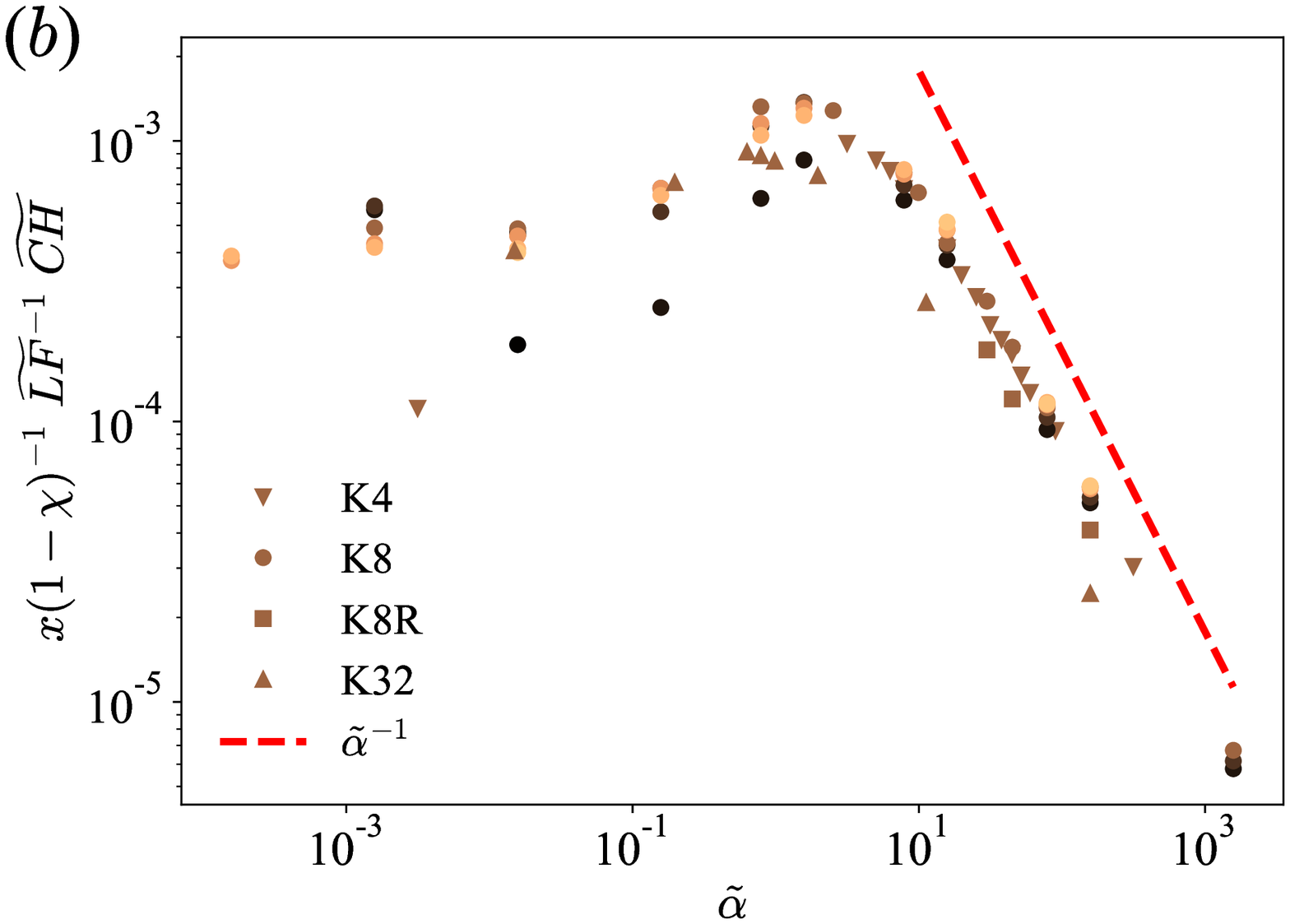}
	}
	\caption{(Colour online) (\textit{a}) Nondimensional total energy versus the collision strength $\talpha$, rescaled by the energy injection rate of the neutrals over the ionization fraction. The red dashed line shows the prediction from equation (\ref{eq:en_predic}), valid in the $\talpha \ll 1$ limit.  (\textit{b}) Nondimensional collisional heating, rescaled by the square of the averaged Lorentz force, versus collision strength $\talpha$. The red dashed line shows the prediction from equation (\ref{eq:ch_predic}), valid in the $\talpha \gg 1$ limit.}
	\label{fig:global}
\end{figure}

Although our predictions seem to qualitatively agree given the snapshots in figure \ref{fig:flow}, we now aim to confirm our results from subsection \ref{subsec:limits} in a quantitative way. In figure \ref{fig:global} we see two subfigures comparing global (time and volume averaged) quantities, where each data point is a single simulation. All runs performed in this study are included. Each subfigure aims to test the predictions made for each extreme of $\talpha$.
Subfigure \ref{fig:global}(\textit{a}) compares total energy with the collision strength rescaled by the energy injection rate of the neutral species over the ionization fraction, based on the right-hand-side of equation (\ref{eq:en_predic}), and with tildes representing suitable nondimensionalization by a combination of $u_f$ and $k_f$. The red dashed line shows the prediction from equation (\ref{eq:en_predic}), which is valid for small values of $\talpha$, and represents the balance between collisional heating and energy injection rate into the neutral species. The collapse of the data -- for various values of ionization fraction, $Re$, and forcing wavenumber -- on a single line, whose slope agrees with the predicted relationship over various orders of $\talpha$, confirms our claims. Subfigure \ref{fig:global}(\textit{b}) focuses on the highly coupled regime, where the prediction for collisional heating was based on an expansion over $\talpha^{-1}$ revealing that $CH$ was given by the square of the difference in the forces acting on each species -- given by equation (\ref{eq:ch_predic}). This prediction is quite general but simplifies significantly for our simulations due to the fact that densities are uniform, $\nu_i = \nu_n$, and $\bv{F}_i = \bv{F}_n$. After nondimensionalizing using $u_f$ and $k_f$ we are left with $\widetilde{CH} = (1-\chi)\chi^{-1} \talpha^{-1} \widetilde{LF}$, where $\widetilde{LF} = \langle | \tilde{\bv{J}} \times \tilde{\bv{B}} |^2 \rangle$, and $\widetilde{(\cdot)}$ denotes the dimensionless version. Moving everything except $\talpha$ to the left hand side we get that, in the high collisional limit, the rescaled collisional heating ($y$-axis of subfigure \ref{fig:global}(\textit{b})) should be proportional to $\talpha^{-1}$, denoted by a red dashed line. We see that indeed the data collapses on a single line, once again for various ionization fractions, $Re$, and forcing wavenumbers. The slope of the rescaled collisional heating seems to approach the theoretical one for large values of $\talpha$, confirming our predictions for the highly collisional limit.

\subsection{Spatial}\label{subsec:spatial}
Our spatially averaged (`global') predictions for the extreme limits of $\talpha$, taken from subsection \ref{subsec:limits}, have been confirmed. However, turbulence is an out-of-equilibrium, multi-scale process whose scale-by-scale analysis can reveal further interesting phenomena that are difficult to identify or predict otherwise. This is the purpose of the current subsection.

In our scale-by-scale analysis we will look at two general quantities -- spectra, which tell how a quadratic quantity is distributed over scales, and fluxes, which tell us how that quadratic quantity is flowing through scales \citep{Alexakis_Review}.
The one-dimensional spectrum $KE_s(k)$ of the kinetic energy of a species $s$ is
\begin{equation}\label{eq:spec_KE}
	KE_s(k) = \frac{1}{2} \sum_{|\bv{k}|=k} |\widehat{\bv{v}_s}|^2(\bv{k}),
\end{equation}
where $\widehat{(\cdot)}$ denotes the Fourier transform of $\bv{\bv{v}_s}$. We will also be looking at the dimensionless collisional heating spectra,
\begin{equation}\label{eq:spec_CH}
	\widetilde{CH}(k) = \frac{(1-\chi)\,\chi\,\alpha}{u^3_f \,k_f} \sum_{|\bv{k}|=k} |\widehat{\vi - \vn}|^2(\bv{k}),
\end{equation}
and the dimensionless spectra of the square of the Lorentz force,
\begin{equation}\label{eq:spec_B}
	\widetilde{LF}(k) = \frac{1}{k_f^2 u_f^4} \sum_{|\bv{k}|=k} |\widehat{\bv{J} \times \bv{B}}|^2(\bv{k}).
\end{equation}

\begin{figure}
	\centering{
	\includegraphics[width=0.32\textwidth]{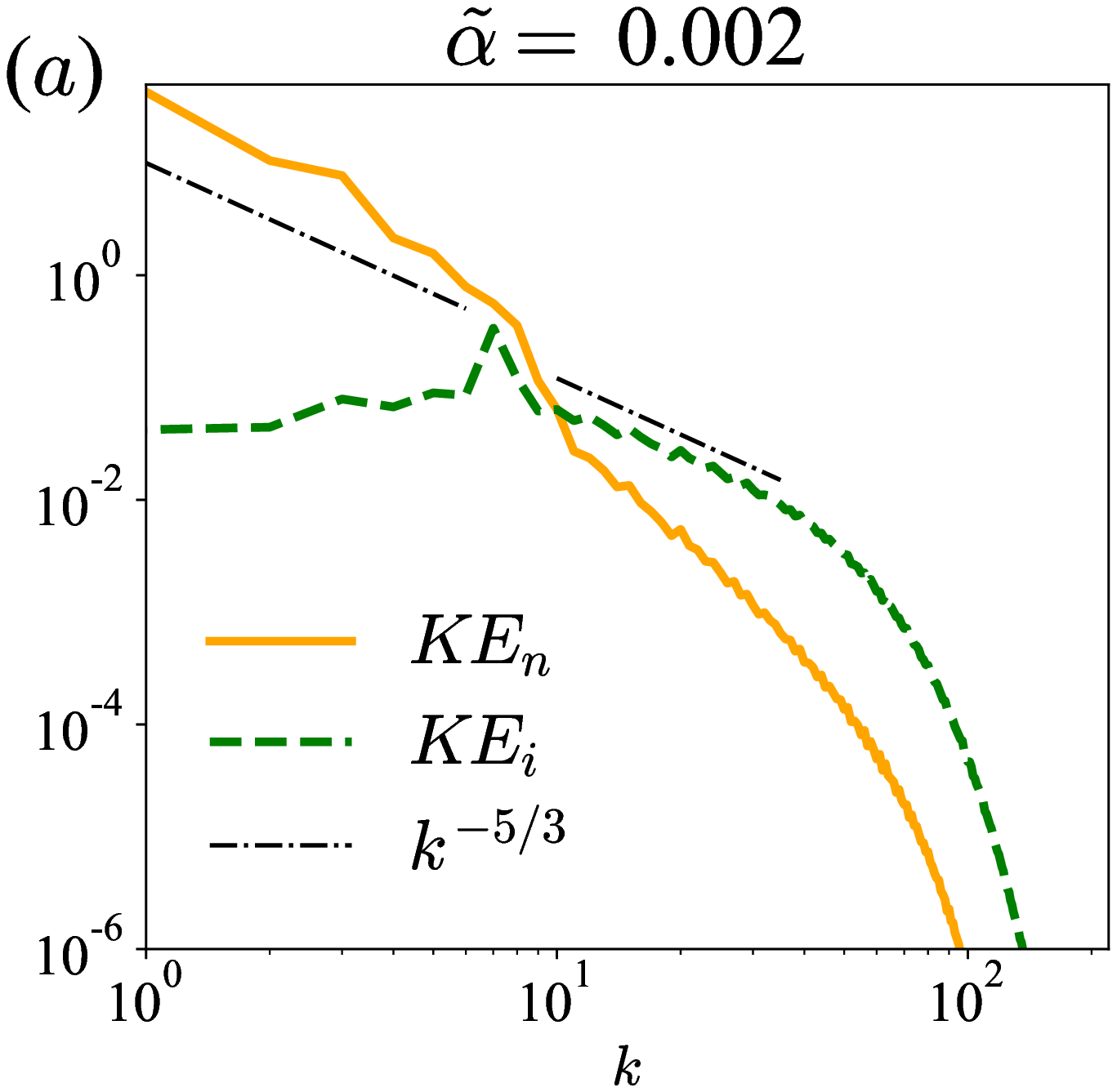}
    \includegraphics[width=0.32\textwidth]{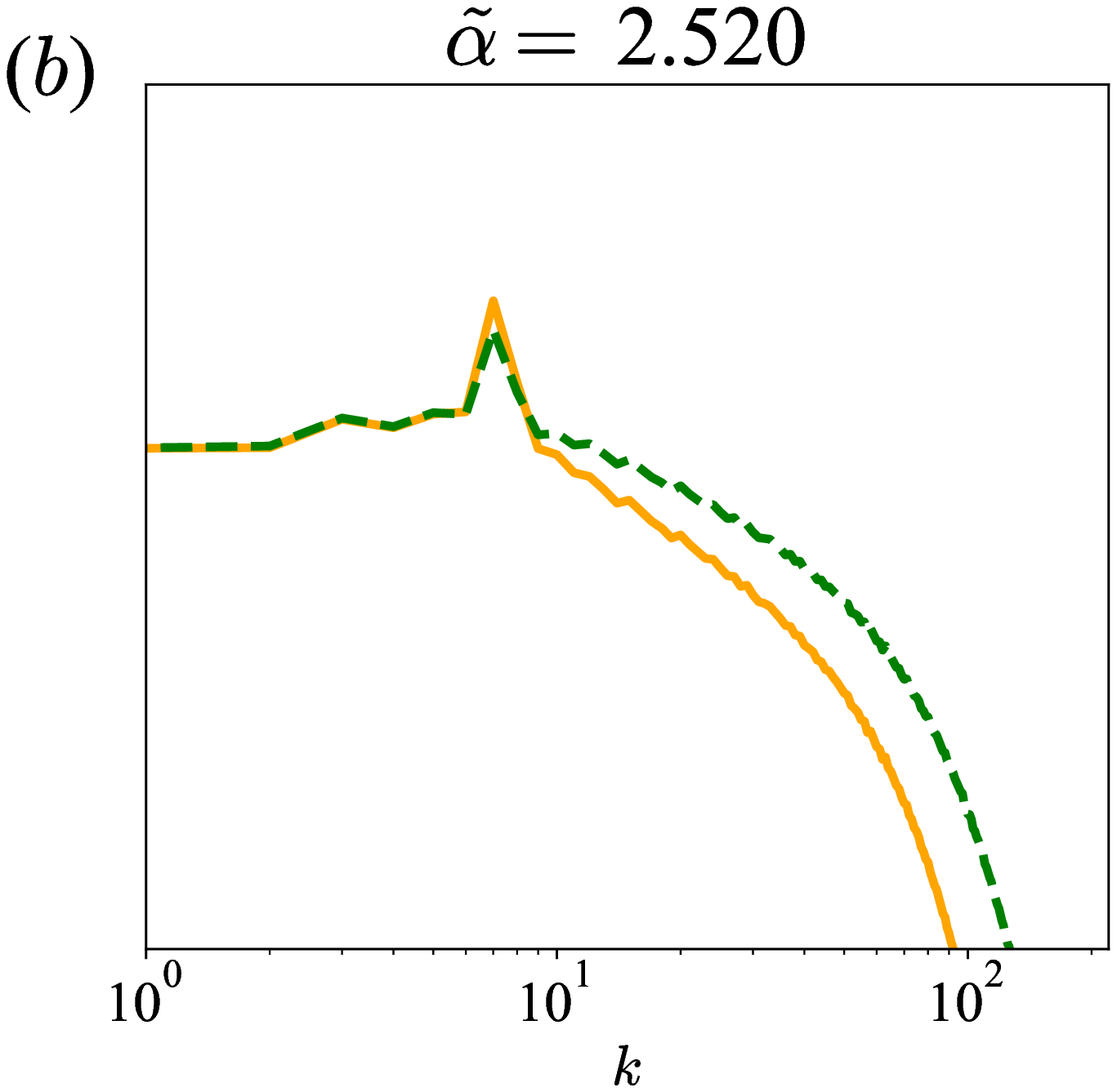}
	\includegraphics[width=0.32\textwidth]{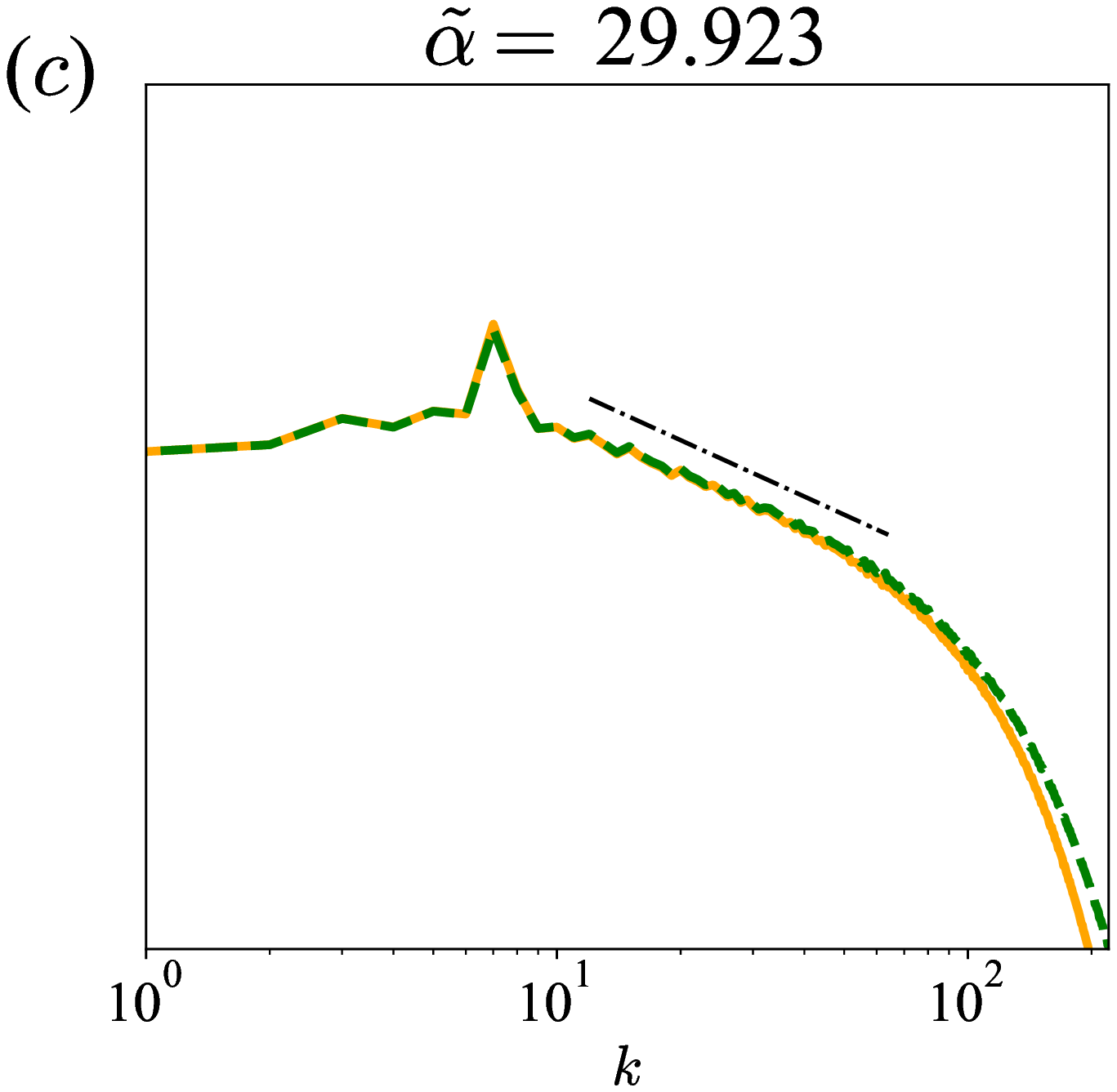}
}
	\caption{(Colour online) Kinetic energy spectra for the neutral species (orange, solid line) and ion species (green, dashed line) at (\textit{a}) $\talpha = 0.002$,  (\textit{b}) $\talpha = 2.520$, and (\textit{c}) $\talpha = 29.923$, all with ionization fraction $\chi=0.5$. As $\talpha$ increases, the two species become more and more coupled, starting from the largest scales which couple first.}
	\label{fig:ke_spec}
\end{figure}

We have seen that $\talpha$ measures the relative strength between collisions and the nonlinear term at a typical scale $L$ with velocity $U$ (or, in the case of our simulations, $k^{-1}_f$ and $u_f$). Therefore, it measures the degree of coupling between the two fluids \textit{at $L$}. However, due to the multi-timescale nature of turbulence, different length scales couple at different values of $\talpha$. The global results from the last subsection showed us how to predict the average coupling of the fluid, given some information about the collision strength at the forcing scale. However, a scale-by-scale analysis allows us to see exactly how the two fluids gradually couple together among scales and to better understand non-extreme cases for $\talpha$.
In figure \ref{fig:ke_spec} we see the steady-state average kinetic energy spectra of both species, for three different values of $\talpha$, as in figure \ref{fig:flow}. These spectra have ionization fraction $\chi = 0.5$ and were taken from the K8 set of runs, forced at intermediate wavenumbers so as to be able to resolve approximate inertial ranges in both large and smaller scales.
In subfigure \ref{fig:ke_spec}(\textit{a}) we have the uncoupled limit, evident by the two spectra and their distinct shapes. The black dot-dashed lines show a $-5/3$ slope, the prediction for the large scales of $KE_n$ and the small scales of $KE_i$. The slope of the neutral kinetic energy is a bit steeper at large scales due to the presence of a condensate. Despite this, as well as a poorly resolved inertial range for the ion kinetic energy, we see reasonable agreement between expected and observed behavior for each individual species. Subfigure \ref{fig:ke_spec}(\textit{b}) shows $\talpha = 2.52$, a moderately coupled case. In this subfigure we see that, indeed, the energies at scales down to the forcing scale seem to be lying on top of each other, implying a coupling at those scales. The collisions are strong enough at these scales to completely remove the inverse cascade of kinetic energy in the neutral species, either by dissipation or transfer of energy to the ions (to be discussed further when we look at the fluxes). However, at scales smaller than the forcing there is a clear distinction between the two energies, implying a lack of coupling. It is this remaining `slippage' between the ions and the neutrals in the small scales, along with the fact that $\talpha \sim \mathcal{O}(1)$, that maintains a large collisional heating, despite the small value of $|\vi-\vn|$ at large scales. In fact, collisional heating is largest at these values of $\talpha$, as seen in subfigure \ref{fig:global}(\textit{b}). Finally, subfigure \ref{fig:ke_spec}(\textit{c}) shows a higher collisional case with $\talpha = 29.923$. In this subfigure we observe that scales smaller than the forcing scale are now beginning to couple, however $\talpha$ is small enough so that the smallest scales are still not coupled. The two fluids look almost identical, like the same MHD fluid. 
At this point a natural question arises: at what scales do the fluids decouple and how does that depend on $\talpha$? We have attempted to numerically investigate the dependence of the decoupling wavenumber, which we are calling $k_{coll}$, on $\talpha$, but were not able to reach a definitive conclusion. Possible issues include: a small inertial range (limited by the numerical capabilities), a failure of time-scale assumptions common in homogeneous, isotropic turbulence which only are valid in a true inertial range (here, dissipation due to collisions might render that irrelevant), and more. We do, however, think that the results are worth showing, and so we have included this work in the appendix \ref{app:k_coll}.

\begin{figure}
	\centering{
		\includegraphics[width=0.32\textwidth]{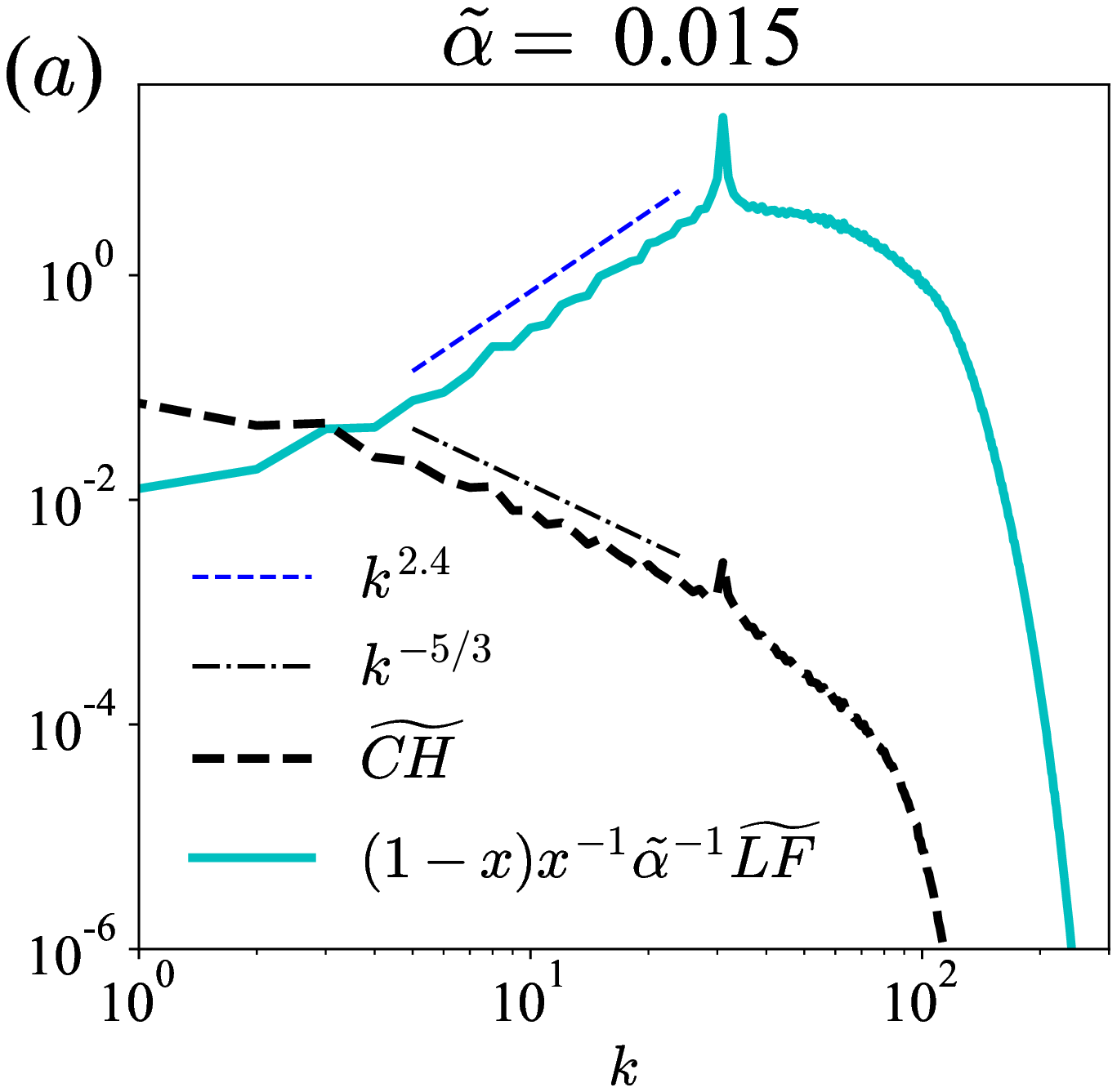}
		\includegraphics[width=0.32\textwidth]{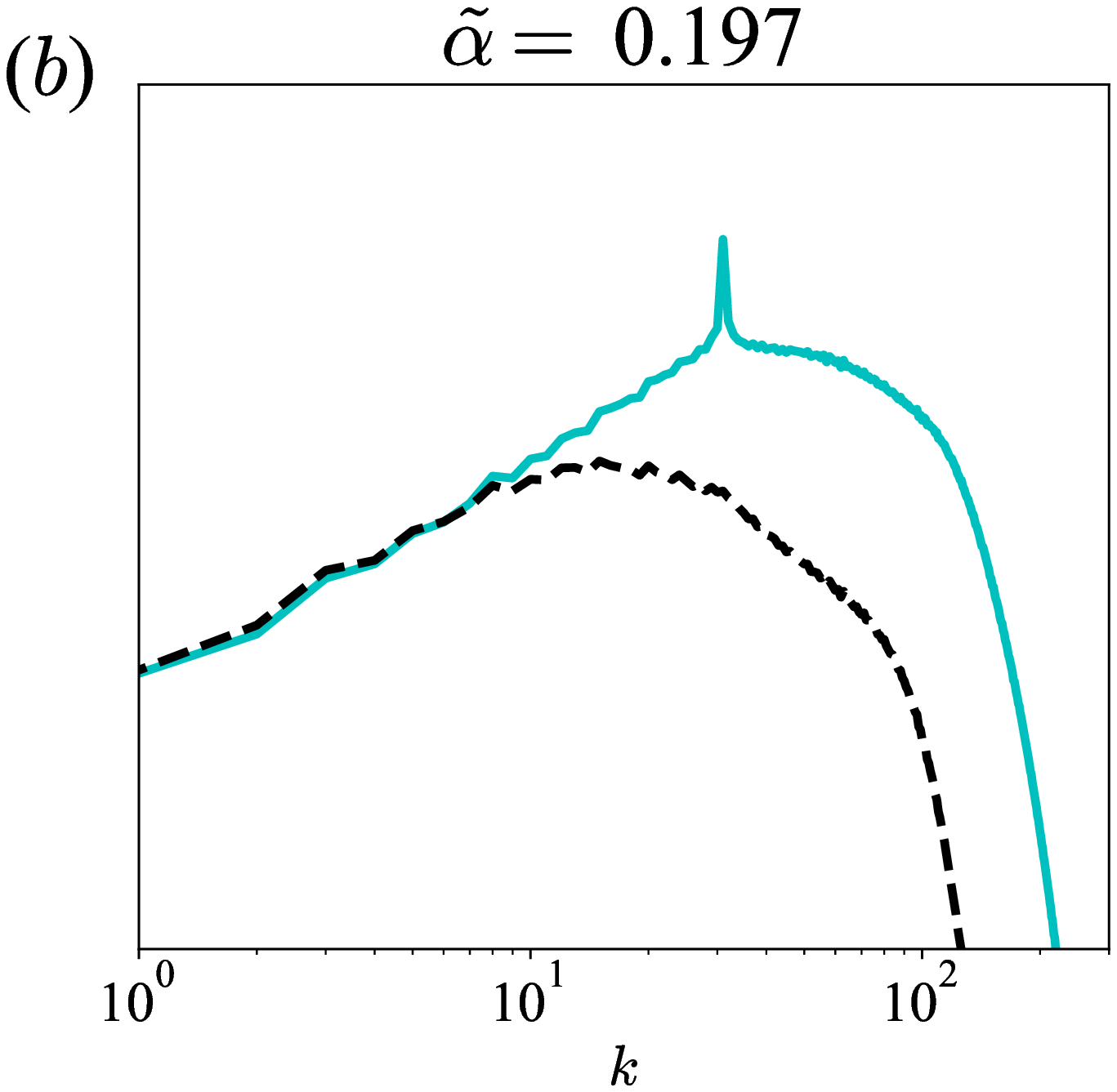}
		\includegraphics[width=0.32\textwidth]{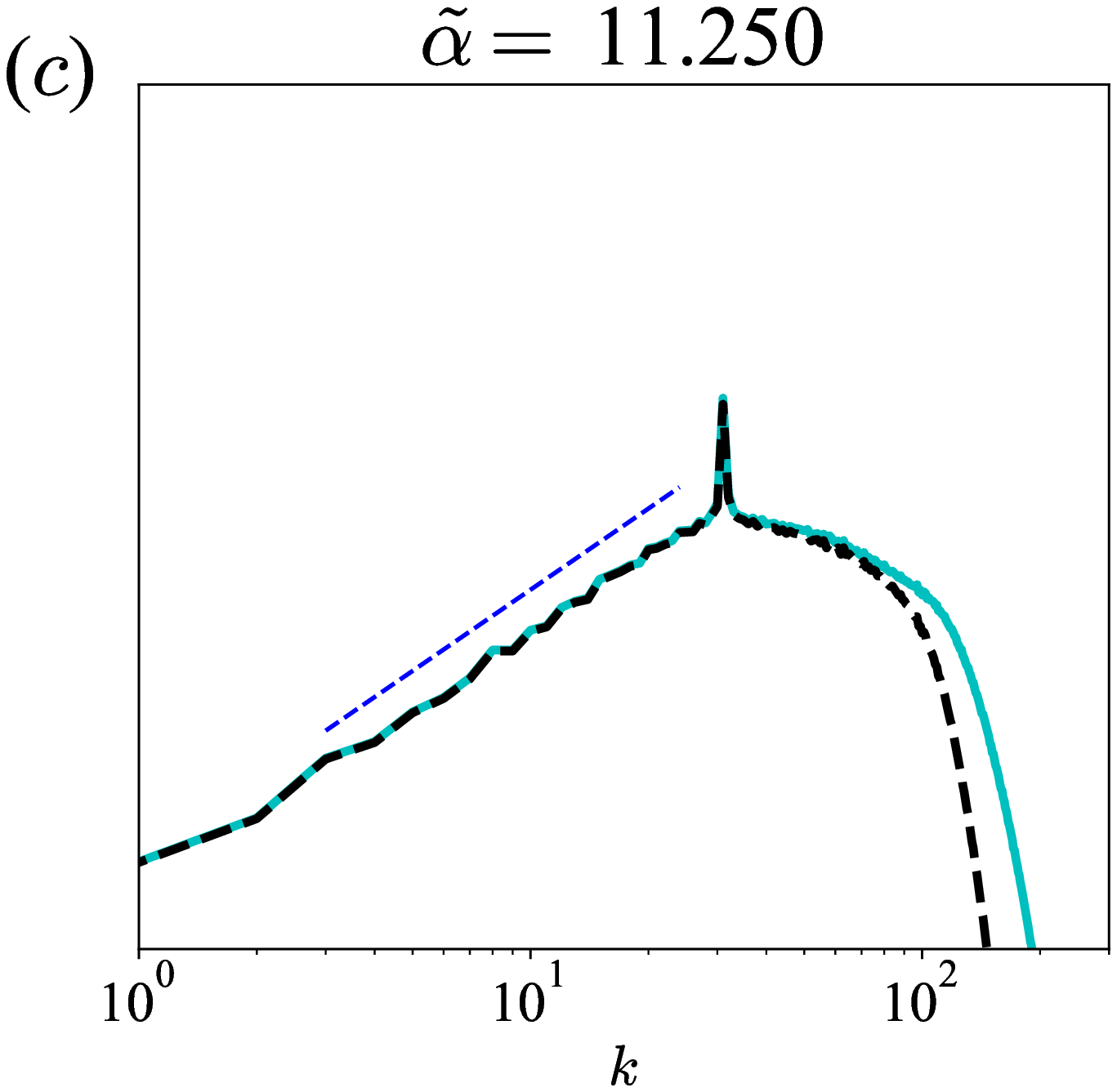}
	}
	\caption{(Colour online) The spectra of dimensionless collisional heating $\widetilde{CH}$ (bold, black, dashed lines) and of the dimensionless right hand side of equation (\ref{eq:ch_predic}), proportional to the spectrum of the Lorenz force squared, $\widetilde{LF}$, (bold, light blue, solid lines). These two curves are expected to be equal when the two species are highly coupled. Shown are three limits of $\talpha$, weak coupling (\textit{a}), moderate coupling,  (\textit{b}), and strong coupling (\textit{c}), allowing one to see the gradual coupling of the species and the change in form of the $CH$ spectrum.}
	\label{fig:CH_spec}
\end{figure}

A quantity intimately related to the degree of coupling between the two species is the collisional heating, which is proportional to $|\vi - \vn|^2$. A scale by scale decomposition of $CH$ can tell us where kinetic energy dissipation (heating) due to collisions is most active, which might be of interest in astrophysical applications. In figure \ref{fig:CH_spec} we plot the nondimensional spectra of $CH$, the black dashed curves, for the usual three cases of $\talpha$. These are taken from steady-state averages in the K32 set of runs, which were forced at small scales.
Focusing only on the black dashed curves for now, we look first at subfigure \ref{fig:CH_spec}(\textit{a}), the low coupling case. Since the two species are not coupled, we can approximately say that at scales larger than the forcing the kinetic energy of the neutrals dominates and hence $CH \propto |\vi-\vn|^2 \approx |\vn|^2 \propto KE_n$. This is confirmed by the $-5/3$ slope shown by the thin dot dashed black line. Hence, the collisional heating is acting like a friction term for the neutral species, causing significant dissipation at the largest scales where the condensate lies. 
Although not shown here, apart from directly dissipating energy to heat, collisions also act to transfer energy from neutrals to ions at large scales (particularly for low ionization fractions).
This possibility is evident by looking at equation (\ref{eq:en_s}) -- when $|\vn|\gg |\vi|$ the sign indefinite term will dominate $|\vi|^2$ for the ion species kinetic energy equation, thus providing the possibility to gain kinetic energy via collisions with the highly energetic neutrals at those scales. 
The species first begin to couple at the largest scales, as we saw when looking at the kinetic energy spectra. This coupling reduces $CH$, causing the length scale of maximal collisional heating to become smaller and smaller as we increase $\talpha$. This is exemplified in subfigure \ref{fig:CH_spec}(\textit{b}), where we already see a significant change in the shape of the collisional heating spectrum, its peak around $k = 15$. This happens until $\talpha > 1$, after which the maximal heating is at the forcing scale and the shape of the $CH$ spectrum remains practically unchanged, with only its magnitude being reduced (proportional to $\talpha^{-1}$). The shape of the collisional heating spectrum at scales that are already coupled is set by the spectrum of $\la |\bv{J} \times \bv{B}|^2 \ra$, as predicted by equation (\ref{eq:ch_predic}). The light blue, solid lines in figure \ref{fig:CH_spec} represent the spectra of the right hand side of this equation, and indeed the two curves lie on top of each other at the scales which are coupled. Thus, understanding the spectrum of the square of the Lorentz force is key to understanding the scale-by-scale collisional heating. The small dark blue dashed line depicting a slope of $2.4$ (found by fitting that region of the spectrum) can be seen in subfigures \ref{fig:CH_spec}(\textit{a}) and \ref{fig:CH_spec}(\textit{c}), showing that the spectrum of the Lorentz force has not changed and is not affected by the collisions. One would still like to know what sets the spectrum of the Lorentz force (and hence of $CH$). A very clear power law with a positive exponent is seen in the figure, so one is tempted to speculate about its origins, particularly using equilibrium spectra arguments, as is done in three-dimensional homogeneous and isotropic turbulence. However, such equilibrium spectra are not necessarily universal if the forcing in the spectral shell around $k_f$ is dense, as is ours  \citep{Alexakis2019}. A prediction of the spectral slope of the Lorentz force, although interesting and relevant for our applications, is beyond the scope of this paper.

\begin{figure}
	\centering{
		\includegraphics[width=0.3\textwidth]{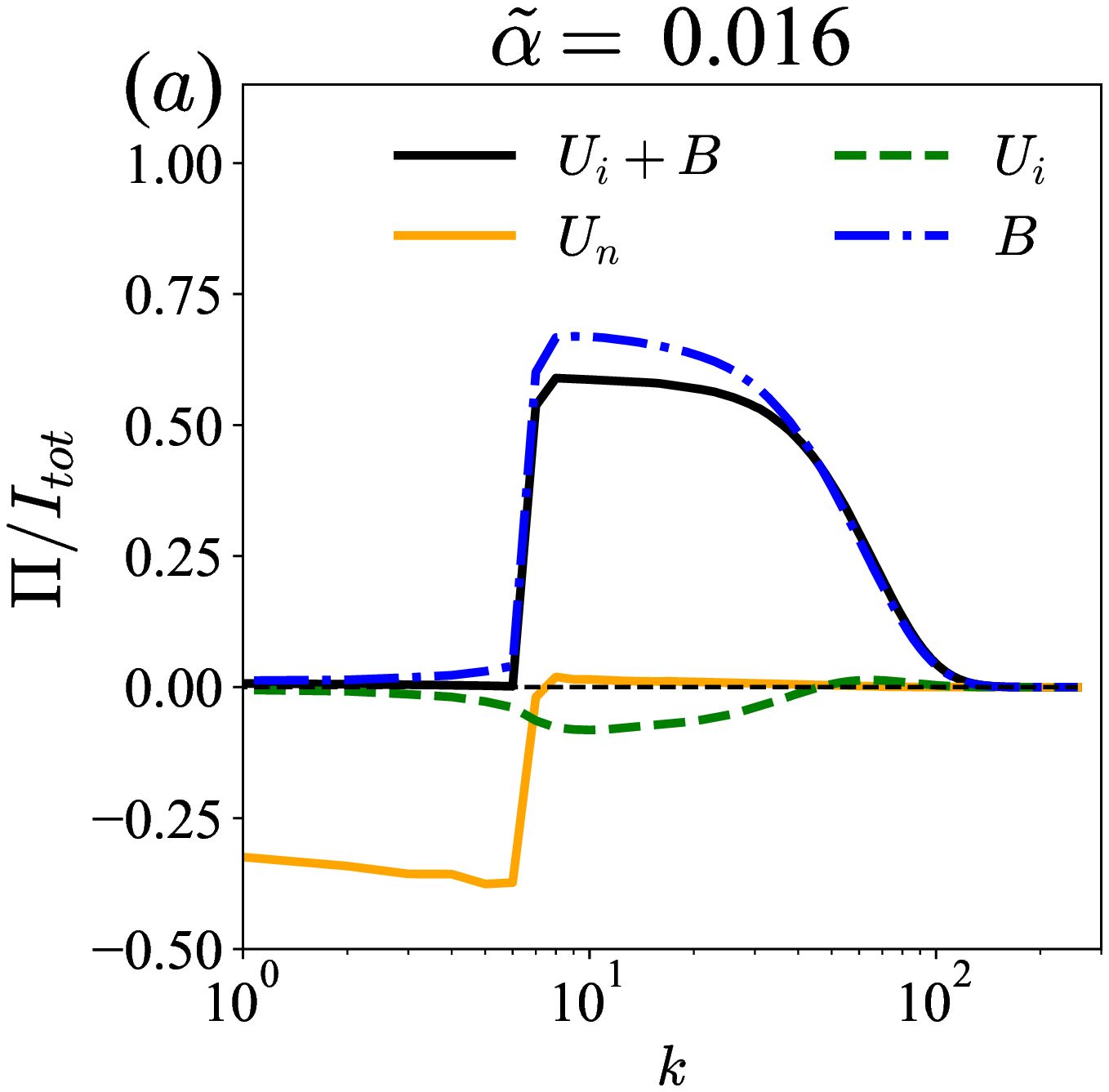}
		\includegraphics[width=0.3\textwidth]{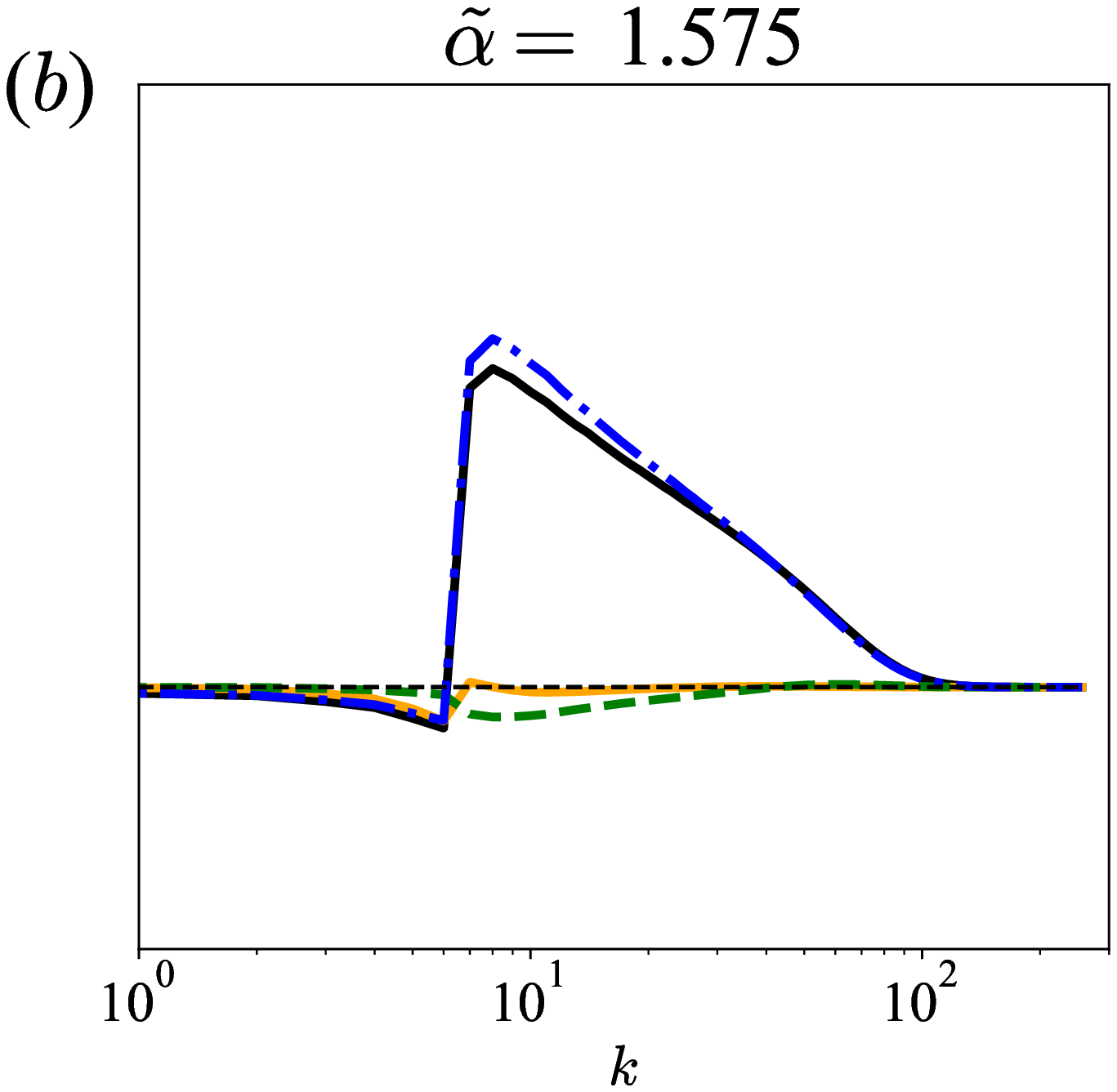}
		\includegraphics[width=0.3\textwidth]{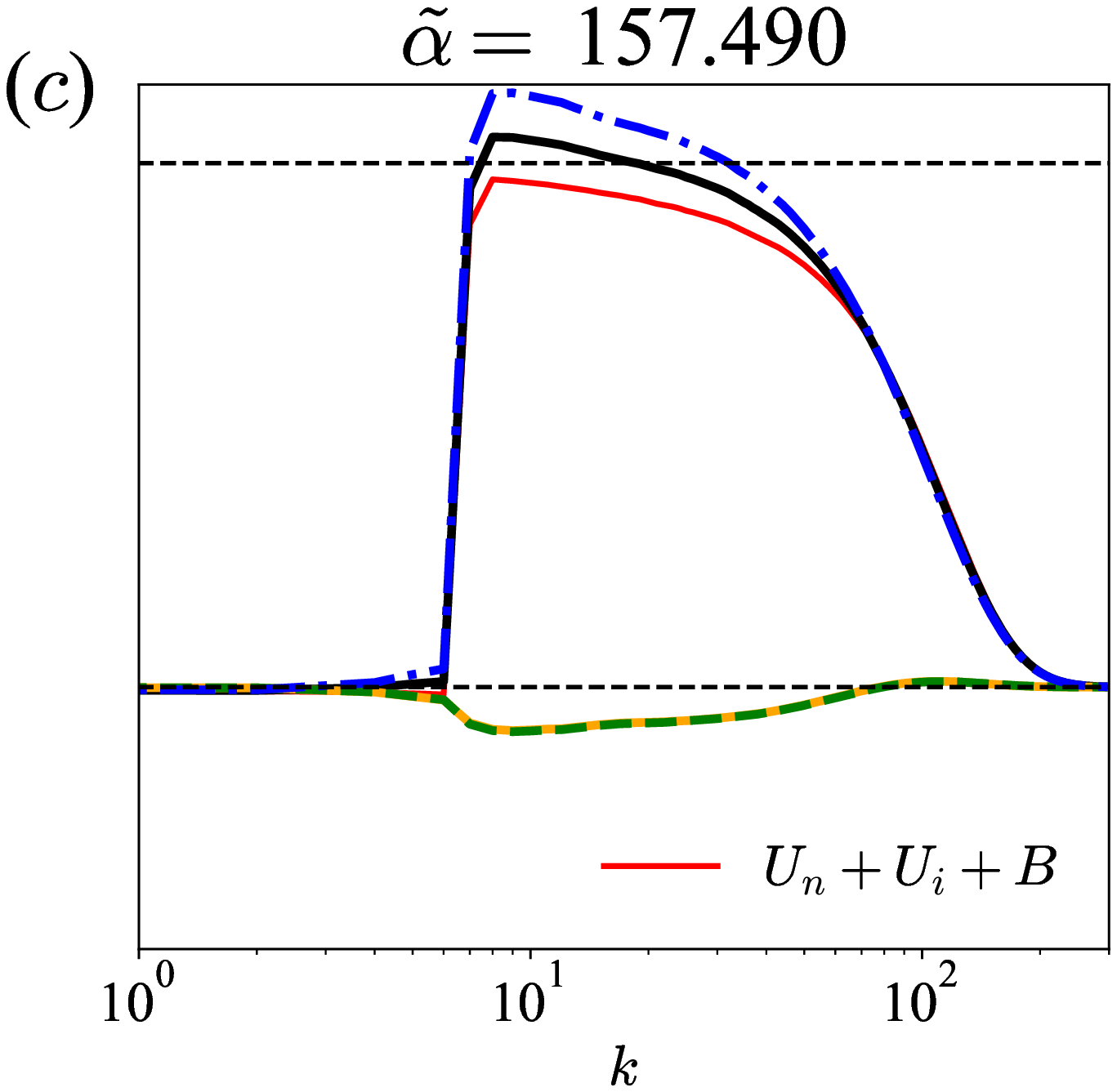}
	}
	\caption{(Colour online) Kinetic component of energy flux for the neutral species, $\Pi_{U_n}$ (orange, solid line), and ion species, $\Pi_{U_i}$ (green, dashed line), as well as the magnetic component of energy flux for the ions, $\Pi_B$ (blue, dot-dashed line). Total ion energy flux is denoted by the solid, black line, and grand total energy flux $(\Pi_{U_n}+\Pi_{U_i}+\Pi_{B})$ by the thin, solid, red line, seen only in subfigure (\textit{c}). The three subfigures depict the three limiting cases for $\talpha$: (\textit{a}) uncoupled, where we see the HD and MHD predictions for fluxes. (\textit{b}) Moderate coupling, which is highly dissipative. (\textit{c}) High coupling, in which the neutral are fully coupled to the ions and we recover the prediction from subsection \ref{subsec:limits} that the system behaves as a single MHD fluid with $\bv{F} = \bv{F}_i + \bv{F}_n$.
}
	\label{fig:fluxes}
\end{figure}

The spectra have revealed the length-scales at which coupling between the species occurs, as well as allowed us to observe at what scales collisional heating dominates and confirm our predictions about the spectral shape of $CH$. While these spectra have informed us of the distribution of various quantities among length-scales, also of interest in turbulence is how certain conserved quantities, such as energy, \textit{move} across scales. This is measured by the (spectral) flux. In this study, we will only look at two components of energy flux, what we call the kinetic component of energy flux for a species $s$, $\Pi_{U_s}$, and the magnetic component of energy flux for the ions, denoted by $\Pi_{B}$. The former is defined to be:
\begin{equation}
\Pi_{U_s}(k) = \la \bv{v}^{<k}_s \cdot \left(\bv{v}_s \cdot \nabla \bv{v}_s \right) \ra,
\end{equation}
where $\bv{v}^{<k}_s$ stands for a filtering of the velocity $\bv{v}_s$ in Fourier space so that only the wavenumbers with modulus smaller than $k$ are kept. The flux $\Pi_{U_s}$ expresses the rate at which kinetic energy of a species $s$ is flowing out of scales larger than $\ell = 2 \pi / k$ due to nonlinear interactions only in velocity. Therefore, if energy is going from large to small scales, the energy flux will be positive, and vice versa. The magnetic component of energy flux for the ion species is defined to be:
\begin{equation}
\Pi_{B}(k) = -\la \vi^{<k} \cdot \left(\bv{B} \cdot \nabla \bv{B} \right) \ra + \la \bv{B}^{<k} \cdot \left( \vi \cdot \nabla \bv{B} - \bv{B} \cdot \nabla \vi \right) \ra.
\end{equation}
As a reminder, in two-dimensional HD turbulence energy flows to larger scales whereas 2D MHD turbulence has been shown to cascade total energy forward to smaller scales. Thus, in an uncoupled system with $\talpha = 0$, we expect the neutral kinetic energy flux $\Pi_{U_n}$ to be negative at scales larger than the forcing and zero at scales smaller than the forcing. On the other hand, $\Pi_{U_i} + \Pi_{B}$ ought to be zero at large scales and positive at scales smaller than the forcing. Although $KE_i$ and $E_B$ aren't individually conserved, the nonlinear interactions which cause the respective fluxes $\Pi_{U_i}$ and $\Pi_{B}$ each conserve energy and can thus also be analyzed. In figure \ref{fig:fluxes} we look at all four fluxes, $\Pi_{U_n}$, $\Pi_{U_i}$, $\Pi_{B}$, and $\Pi_{U_i}+\Pi_{B}$, normalized by the total energy injection rate $I_{tot} = I_n + I_i + I_B$. We look at three cases of $\talpha$, the first two taken from the K8 set of runs, and the large $\talpha$ limit from the K8R run (all with $\chi = 0.5$). These fluxes are averaged over the steady state.
The $\talpha \ll 1$ case, subfigure \ref{fig:fluxes}(\textit{a}), confirms the uncoupled predictions. The neutral kinetic energy flux (orange solid line) is negative and roughly equal to the neutral injection rate: $I_n/ I_{tot} = 4/9$ (recall that $I_n = I_i = 4 I_B$). The total ion energy flux (black solid line) is positive and seen to be close to $(I_i+I_B)/I_{tot} = 5/9$. Since $\talpha \neq 0$, we do see a hint of some large scale energy dissipation in the neutrals, given by the slight negative slope of the flux at large wavenumbers. Turning now to the case where $\talpha \approx 1$, subfigure \ref{fig:fluxes}(\textit{b}) shows that the neutral kinetic energy flux is practically zero for all wavenumbers, meaning that collisions have almost completely stopped any nonlinear transfer of kinetic energy among the neutrals. At the forcing scale and larger, the two species are coupled and behave like MHD, as we have seen. Since MHD has no inverse cascade of kinetic energy, the neutrals have no inverse cascade at these scales. At scales smaller than the forcing the species are not yet coupled, and since the neutral species does not have a forward cascade of energy, the flux remains zero. However, there is still constant injection rate of kinetic energy into the neutrals of magnitude $I_n$, and this energy must go somewhere. That energy is either dissipated away by collisions, or transferred to the ion species. Seeing as, at best, about half of the total energy injection rate is being transferred spectrally by the ions, it seems that in this case about half of the energy injected at the forcing scale is immediately dissipated by collisions, the rest being transferred away by the ion species. For the case of uncoupled 2D MHD turbulence, one would expect this flux of energy by the ion species and magnetic field to be constant in the inertial range down to the dissipation scales. However, in subfigure \ref{fig:fluxes}(\textit{b}) we see a gradual drop off of energy flux. This non-constant flux is a sign of strong dissipation of energy in the would-be inertial range due to collisional heating, which is largest in the runs where $\talpha \sim \mathcal{O}(1)$, as we have seen.
Moving on to subfigure \ref{fig:fluxes}(\textit{c}), we notice that the ion and neutral species are fully coupled, even at the small scales, as seen by the fact that $\Pi_{U_n} = \Pi_{U_i}$. Now that the two species are coupled throughout most of the scales, the collisional heating is very weak and does not dissipate practically any energy that is being injected at the forcing scale. This subfigure indicate that \textit{all} of the energy being injected by the forcing -- including the neutral energy injection -- is going to small scales. This is shown by the total energy flux from all fields, denoted by the thin, red, solid line, which we see is positive and very close to 1. This is consistent with our $\talpha \rightarrow \infty$ limit studied in subsection \ref{subsec:limits}, where we showed that the system behaves as a single MHD fluid with $\bv{F} = \bv{F}_i + \bv{F}_n$, as seen in equation (\ref{eq:u}).
Sticking to $\vi$ and $\vn$, what seems to be happening is that the two species exchange energy between each other very efficiently due to collisions, but, since ions also exchange energy with the magnetic field, part of the energy being injected into the neutral species ends up as magnetic energy, which is transferred to smaller scales.
Although it is not shown here, we ran a few decaying turbulence experiments with no forcing to see if this description holds true. Indeed, after initializing the two species with identical random initial conditions (and $KE_i = KE_n = E_B$ at $t = 0$), the runs with very large $\talpha$ showed larger initial magnetic field growth, implying an indirect transfer of energy from the neutrals.

\section{Discussion \& Conclusions}\label{sec:conclusions}
\subsection{Conclusions}
In this work we have investigated the two-dimensional partially-ionized magnetohydrodynamics (PIMHD) system, a two-fluid model used for studying plasmas where some fraction of the ions have recombined to form neutral molecules which interact with the other ions via collisions. Although ionization fraction certainly plays a role in the dynamics of such plasmas, we focused more on the role that collisions have on the dynamics. In the limit where collision time-scales are long compared to dynamical time-scales, we found that the two species were weakly coupled and behaved like their uncoupled fluid counterparts -- hydrodynamics (HD) for neutral species and magnetohydrodynamics (MHD) for the ions. In this limit, collisions act like a frictional force for whichever species has larger kinetic energy at that scale. Hence, collisions took the role of a large-scale dissipation for the neutral species. Using this observation and an energy balance argument we were able to predict the amount of collisional heating in these runs. For low ionization fractions, it is possible that ions may gain some kinetic energy via collisions at these large scales. At intermediate collision strengths, where dynamic and collision time-scales are similar, we found the dynamics to be quite dissipative. About half of the energy injected into the system is dissipated immediately at the forcing scale, and another quarter of it is dissipated by collisions along the forward cascade range. In this case, the neutral species has little-to-no nonlinear energy transfers. For collision time-scales much shorter than dynamical time-scales we found that the two species are coupled and act like a single MHD fluid whose density, pressure, viscosity, and external forces are the sum of each from the two species. This was confirmed numerically, particularly in subfigure \ref{fig:fluxes}(\textit{c}) which showed all of the energy injected at the forcing scale being transferred to smaller scales. This species coupling in turn reduces the collisional heating significantly, which we showed, and also confirmed numerically, is determined by the square of the difference of accelerations of each species. In our specific implementation, the nature of the Lorentz force meant that the collisional heating no longer dominates at the largest scales (as in the low collisional case) but is present at smaller scales. A scale-by-scale analysis of our runs allowed us to further understand how the coupling of the two species gradually occurs as the collision time-scale decreases. Although we numerically investigated two-dimensional PIMHD, our results for the strong collisional case are expected to hold for the three-dimensional counterpart since dimensionality did not play a role in the derivation of the predictions made.

\subsection{Connection to Jupiter}
This work was motivated by the transition region between the ionized interior and neutral upper atmosphere of gas giant planets such as Jupiter, where partial ionization effects are expected to be present. The current treatment of the dynamics in the transition region have been single-fluid models, either of MHD with large resistivity, or HD with a drag term to parametrize any magnetic effects (``MHD drag''). Given our characterization of the behavior of such a two-fluid plasma, we will attempt to apply some of what we learned here to the transition region of gas giant planets. In order to do so we must first estimate the ratio of eddy turnover time to collision time-scale, $\talpha$, for the case of Jupiter's transition region, whose densities, temperatures, pressures, and typical velocities and length-scales are better constrained than any other gas giant planet. Following equation (\ref{eq:talpha}), we must choose a typical velocity $U$ and length-scale $L$, as well as total density $\rho_{tot}$, and finally the collision strength $\alpha$. Typical velocities for the transition region can range from $U = 10^{-2} \ m \, s^{-1}$ close to the edge of the ionized interior, at a radius of roughly $0.9$ Jupiter radii, to $U = 10^2 \ m \, s^{-1}$ at the surface \citep{Kaspi2018}. We take the typical length-scale in this region from \cite{Cao2017}, who estimated a convective length-scale of $L = 6 \times 10^5 \ m$. The total density is found from simulations by \cite{French2012}, who give values of $\rho_{tot} = 10^3 \ kg \, m^{-3}$ at $0.9$ Jupiter radii to $\rho_{tot} = 2 \times 10^2 \ kg \, m^{-3}$ closer to the surface. The final piece is the collision strength $\alpha$, whose value is much harder to constrain, given the fact that to this date no plasma experiment has been able to measure collision cross-sections at such high densities. Therefore, any value used in our estimate will be questionable, but we feel it would be best to give some estimate rather than none. Following section 5.2.1 in \cite{Meier2011} and section III.D.2 from \cite{Meier2012}, under the assumptions that ions and neutrals have the same temperature and that both are in thermal equilibrium, we get the following expression for $\alpha$:
\begin{equation}\label{eq:alpha_calc}
\alpha = \alpha_1 \sqrt{T} - \alpha_2\ln\left(205 \sqrt{T}\right)\sqrt{T},
\end{equation}
where $\alpha_1 = 1.34 \times 10^{11} \ m^2 kg^{-1}s^{-1}K^{-1/2}$ and $\alpha_2 = 8.78 \times 10^9 \ m^2 kg^{-1}s^{-1}K^{-1/2}$.  We should note that these numbers are taken from cross-section estimates for charge exchange reactions (e.g., an electron hopping from a neutral molecule to an ionized molecule), which are expected to be dominant even at the relatively low (sub-$10,000 \ K$) temperatures considered here (E. T. Meier, personal communication), rather than more standard collisions.
However, the functional form of the collision term does not change. To get an estimate for $\alpha$ at 0.9 Jupiter radii and near the surface we use the temperature profile from \cite{French2012}, giving about $5000 \ K$ and $1000 \ K$, respectively. This in turn results in a range of $\alpha$ between $3.4 \times 10^{12} \ m^2 kg^{-1}s^{-1}$ in the interior and $1.8 \times 10^{12} \ m^2 kg^{-1}s^{-1}$ closer to the surface. Previous studies on shock waves in the interstellar medium \citep{Smith1997} and for hot Jupiters \citep{Perna2010}, have gotten similar values. Due to the temperature dependence, and the relatively small size of the transition region in Jupiter, the collision strength does not change much. In a planet such as Saturn, with a much larger transition region \citep{Cao2017}, this might not be the case. Combining everything, we are able to now give an estimate for the range of $\talpha$: closer to 0.9 Jupiter radii we estimate $\talpha = 2 \times 10^{23}$, whereas towards the surface we estimate $\talpha = 2 \times 10^{18}$. Given the $\talpha^{-1}$ dependence of $CH$ from equation (\ref{eq:ch_predic}), this would imply an incredibly small (and presumably negligible) amount of collisional heating in the transition region of Jupiter, as well as single-species MHD-like dynamics, with two-fluid effects being negligible. In fact, $\alpha$ is so large that the two species would be coupled (and therefore behave as an MHD fluid) even for extremely low ionization fractions -- since $\tau_{eddy}\nu_{in} = L \rho_i \alpha U^{-1}$ would be very large. This therefore seems to validate the single-fluid MHD models used so far in dynamo studies which include the transition zone \citep{JONES2011120,Gastine2014,Jones2014,DIETRICH201815}. We want to emphasize, however, that, although the single-species MHD description holds throughout most of the transition region, whether the fluid behaves as an MHD or HD fluid depends on whether the Lorentz force is significant or not. Since we also expect $\eta_{MHD}$ to depend on temperature, it's quite possible that in the transition region the single-species description, although technically an MHD fluid, does not feel the Lorentz force and thus behaves as an HD system. Thus, the transition from MHD to HD could happen continuously, although not because neutrals make up the majority of the fluid, 
but because the Lorentz force becomes weaker and weaker as the magnetic diffusivity significantly increases as a function of radius, leading to neutral-like behavior of the MHD fluid \citep{Tobias2007,Seshasayanan2014,Seshasayanan2016}. As this transition happens we go into the regime of low \textit{magnetic} Reynolds number MHD turbulence, which is where the MHD drag prescription is applied.

\section*{Acknowledgments}
The authors would like to thank Keaton J. Burns, Jeffrey S. Oishi, Basile Gallet, Daniel D. B. Koll, and Eric T. Meier for many helpful discussions throughout the development of the project. This project was funded by a grant from the National Science Foundation (OCE-1459702).

\section*{Declaration of Interests}
The authors report no conflict of interest.

\appendix
\section{Limiting cases of ionization fraction $\chi$}\label{app:x_lims}
Note that no longer having both $\chi \sim \mathcal{O}(1)$ and $(1-\chi) \sim \mathcal{O}(1)$ now means that the two species feel collisions to a different degree, since, as we saw in subsection \ref{subsec:limits}, $\tau_{coll,i} = \tau_{coll}/(1-\chi)$, and $\tau_{coll,n} = \tau_{coll}/\chi$. In the fully ionized limit, $\chi \rightarrow 1$, the collisional time-scale in the ion equation, $\tau_{coll,i}$, becomes very large, yet for neutrals $\tau_{coll,n}$ doesn't change. This results in the ions not feeling collisions in the low collisional limit, whereas neutrals are still affected. In the high collisional limit the collision term is still at least order one for both species, however the highest order dynamics are slightly different. Away from the dissipation range, a similar expansion in $\talpha^{-1}$ results in $\vn^{(0)} = \vi^{(0)} + (\bv{F}_n -\nabla p_n)/ (\rho_i \rho_n \alpha)$ so that the order-one dynamics is simply the MHD of $\vi$ but with a modified pressure $P = p_i + p_n$ and force $\bv{F} = \bv{F}_i+\bv{F}_n$. In the low ionization limit, $\chi \rightarrow 0$, it is now the neutrals that are less affected by collisions. This limit is a bit more involved because it encompasses the breakdown of the validity of other approximations in the derivation of the MHD equations, regardless of the value of $\talpha$. There are two parameters of interest to recall: the electron to ion mass ratio, $\beta \equiv m_e/m_i$, and the ratio of ion skin depth, $d_i \equiv \sqrt{(c^2 m_i)/(4 \pi e^2 n_i)}$, to typical length $L$, which we call $\lambda$. Here $m_e$ and $m_i$ are the masses of the electron and ion, respectively, $c$ is the speed of light, $e$ is the electric charge of the electron, and $n_i = \rho_i/m_i$. In the derivation of MHD, these two parameters are taken to be much smaller than unity. However, when deriving PIMHD, ratios of $\chi$, $\lambda$, and $\beta$ appear and so, when $\chi \ll 1$, terms that would normally be negligible in MHD are no longer small. It can be shown that, if $\chi \sim \lambda \sim \beta \ll 1$, then the induction equation (\ref{eq:fullB}) must be modified to include the Hall term, so that the nonlinear term becomes $\nabla \times (\vi \times \bv{B} - d_i (\bv{J} \times \bv{B})/\rho_i)$ \citep{Pandey2008}. Note that this is now true even at large scales, not just small scales, as is normally the case. In the low collisional limit, the neutrals don't feel the collisions whereas the ions do. On the other hand, when $\talpha \gg 1$, we have, to highest order, $\vi^{(0)} = \vn^{(0)} +(\bv{J} \times \bv{B} - \nabla p_i + \bv{F}_i) / (\rho_i \rho_n \alpha)$, leading to what is typically called ``ambipolar MHD'' in the astrophysics community:
\begin{subequations}
	\begin{equation}
	\rho_{n} \left( \frac{\D}{\D t} + \vn \cdot \nabla\right) \vn = -\nabla (p_n+p_i) +  \bv{J} \times \bv{B} + \bv{F}_i+\bv{F}_n, \label{eq:ambi} 
	\end{equation}
	\begin{equation}
	\frac{\D \bv{B}}{\D t} = \nabla \times \left(\vn \times \bv{B} - \frac{d_i}{\rho_i}\bv{J} \times \bv{B} + \frac{1}{\rho_i \rho_n \alpha} (\bv{J} \times \bv{B} - \nabla p_i + \bv{F}_i) \times \bv{B} \right). \label{eq:ambi2}
	\end{equation}
\end{subequations}
Counter to intuition, despite the neutrals making up the majority of the density in the fluid, the neutrals indirectly interact with the magnetic field via their collisional interaction with the ions and the dynamics is still that of MHD. It is only when $\chi \ll \lambda \sim \beta \ll 1$ that the behavior of the fluid reverts to regular hydrodynamics without interaction with the magnetic field. 
We should note here that we have treated $\chi$ and $\eta_{MHD}$ from equation (\ref{eq:eta}) independently and have effectively held the latter fixed while varying $\chi$. In our claim that it is possible to have MHD behavior even for $\chi \ll 1$ we have assumed that the magnetic diffusivity is low enough so that the Lorentz force is still order one. In a realistic situation, it is likely that ionization fraction and idealized magnetic diffusivity $\eta_{MHD}$ are both related to temperature and so as $\chi \rightarrow 0$ we would also expect $\eta_{MHD}$ to increase significantly; therefore it is possible that the Lorentz force is negligible when $\chi \sim \beta \sim \lambda \ll 1$, making the dynamics that of HD \citep{Tobias2007,Seshasayanan2014,Seshasayanan2016}. However, this is a separate effect than that of ionization fraction, and our claims above still hold in general.

\section{Determining $k_{coll}$ vs. $\talpha$}\label{app:k_coll}
\begin{figure}
	\centering{
		\includegraphics[width=0.49\textwidth]{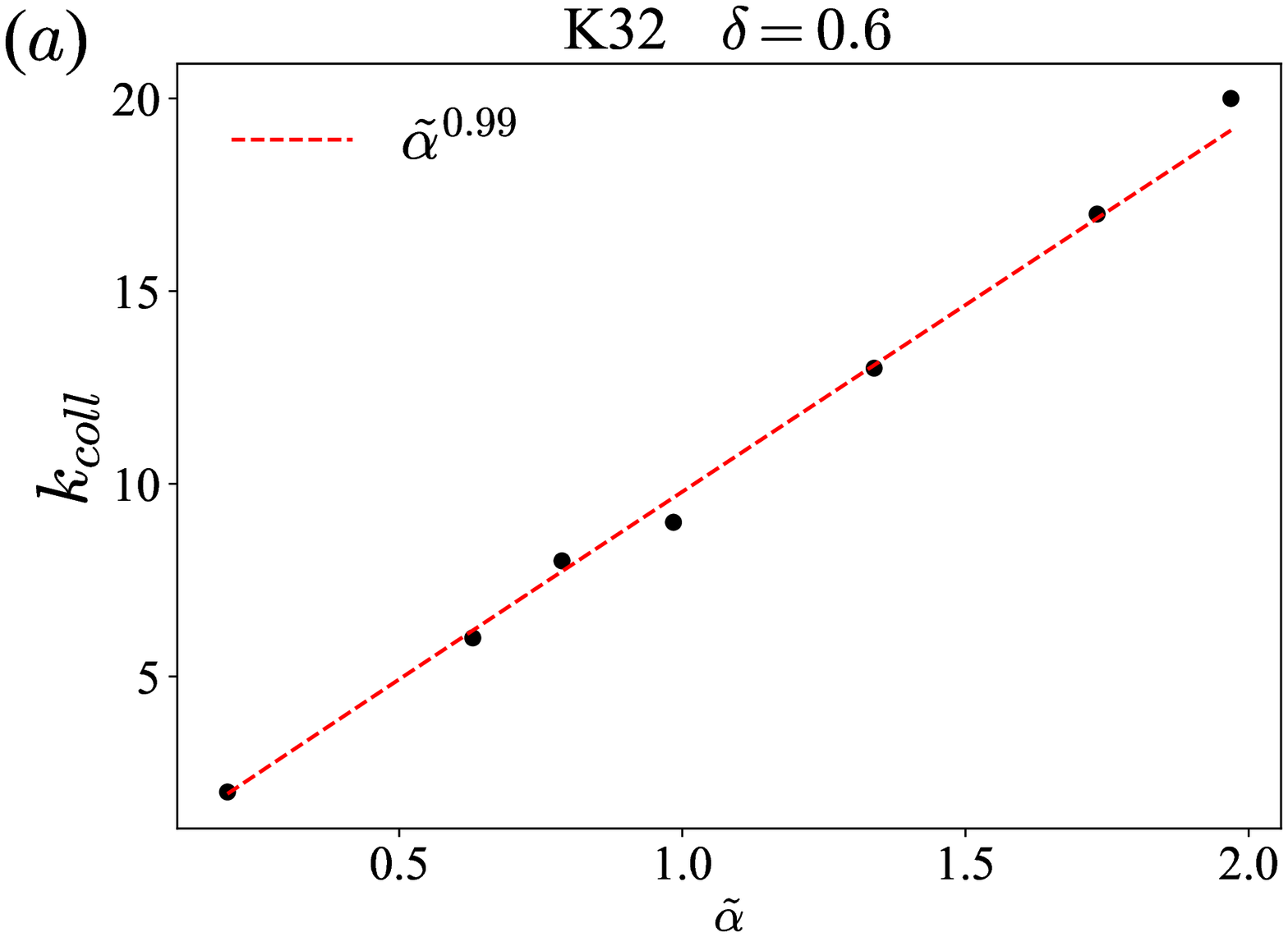}
		\includegraphics[width=0.49\textwidth]{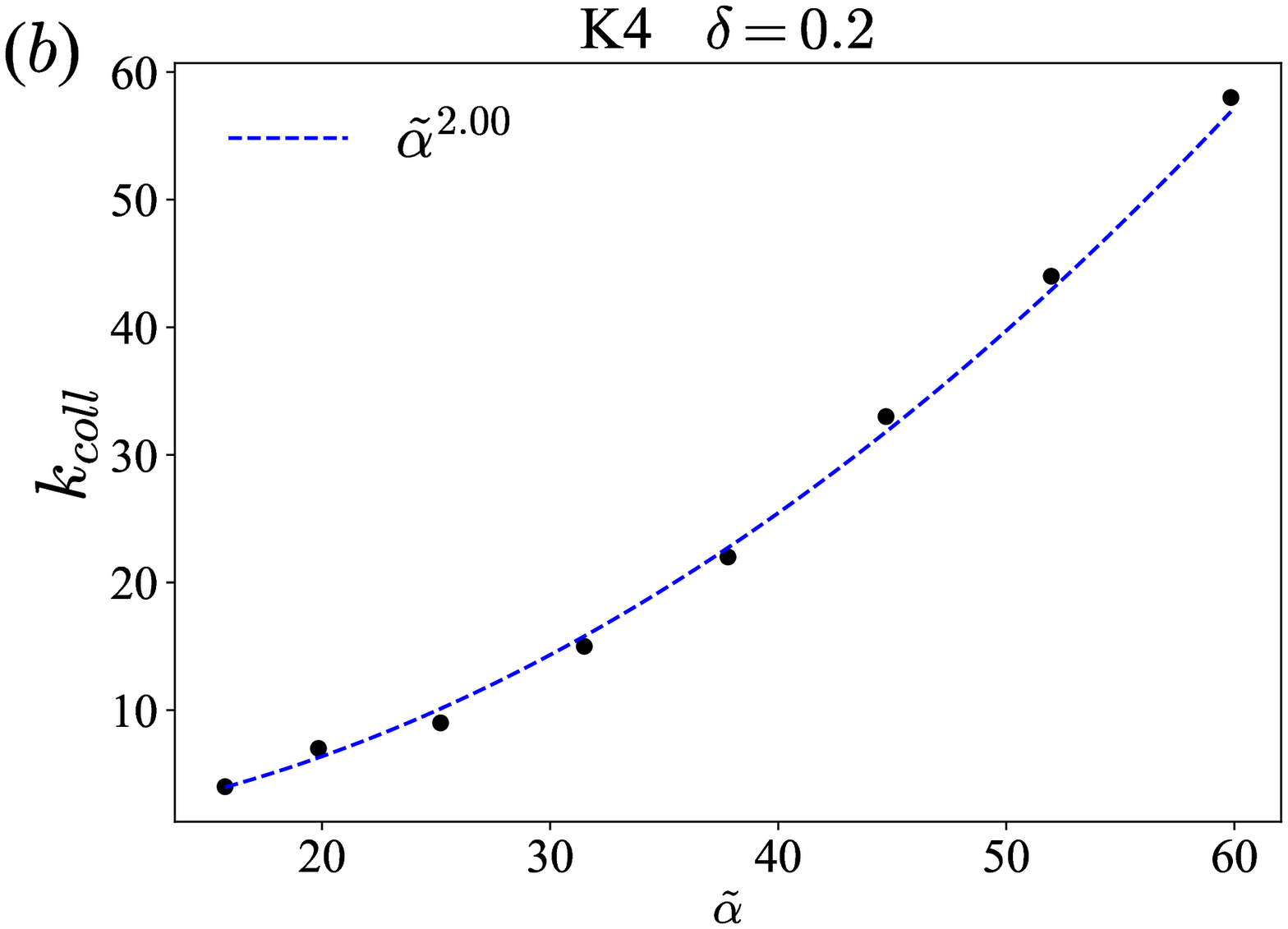}
	}
	\caption{(Colour online) Decoupling wavenumber $k_{coll}$ versus collision strength $\talpha$ for (\textit{a}) $k_{coll}<k_f$, and (\textit{b}) $k_{coll}>k_f$. We define $k_{coll}$ such that $|\bv{D}|(k_{coll})/|\bv{V}|(k_{coll}) = \delta$. The dashed lines represent the best-fit power law, whose exponent is shown in the legend.}
	\label{fig:k_coll}
\end{figure}

A typical argument would tell us that when the eddy turnover time at a scale $k_{coll}^{-1}$, $\tau_{eddy}(k_{coll})$, is equal to the collision time, $\tau_{coll} = (\rho_{tot}\alpha)^{-1}$, then that length-scale is coupled. This sort of argument has been used before for estimating decoupling scales for both MHD waves and nonlinear dynamos in a partially-ionized system \citep{Xu2016,Xu2017}. This time-scale argument relies on an important assumption: that the velocity scale in the nonlinear (or wave) term and collision term have similar orders of magnitude independent of $\talpha$, i.e. $k_{coll} U^2 \sim \rho_{tot} \alpha U$, making $U$ cancel out and leaving one with a ratio of time-scales. This assumption is equivalent to stating that $|\bv{V}| \sim |\bv{D}|$ (defined in subsection \ref{subsec:global}). However, the results of subsection \ref{subsec:limits} tell us that $|\bv{D}| = |\vi-\vn| \propto \alpha^{-1}$, whereas $|\bv{V}| \sim \mathcal{O}(1)$. Although this scaling with $\talpha$ is only true for scales that are already highly coupled, we only expect $|\bv{D}|\sim|\bv{V}|$ far from the coupling scale, thus making the scaling nontrivial. The difficulty arises in the fact that, by definition, at the coupling scale we are not in any extreme limit of $\talpha$ and so we cannot make a general statement about how $|\bv{D}|$ scales with $\talpha$. Furthermore, although it is not shown, we note that the spectra for $|\bv{V}|$ and $|\bv{D}|$ don't have the same power law exponents, implying a different $\tau_{eddy}$ tendencies with $k$, further complicating any length-scale analysis for finding $k_{coll}$.
We therefore revert to simply observing the decoupling wavenumber $k_{coll}$ versus $\talpha$ in our simulations. We define $k_{coll}$ such that $|\bv{D}|(k_{coll})/|\bv{V}|(k_{coll}) = \delta$, where we have the freedom to choose $\delta$ as long as it is small. This is equivalent to saying that the fluid is coupled when $|\vi - \vn| \ll |\chi\vi + (1-\chi)\vn|$, with the freedom to define just how much smaller. In figure \ref{fig:k_coll} we see $k_{coll}$ versus $\talpha$ for two cases: (\textit{a}) $k_{coll}<k_f$, and (\textit{b}) $k_{coll}>k_f$. For each case, a $\delta$ was chosen so that the maximum number of runs could have $k_{coll}$ within the inertial range. Only these runs were kept in the figure.

The former case is taken from various runs in the K32 set, which is forced at small scales to be able to better resolve the inertial range at scales larger than the forcing. The latter is taken from various runs in the K4 set, which is forced at larger scales to be able to better resolve the inertial range at scales smaller than the forcing. Notice that different values of $\delta$ are used for each case, but that $\delta<1$ for both cases. Due to the limited scale-separation possible in numerical experiments, this restricted the possible values of $\delta$ and made it difficult to test the robustness of the calculations of $k_{coll}$ as one varies $\delta$. The dashed lines represent best-fit power laws, whose exponents are seen in the legends of each subfigure. Although the power laws are very close to integer values, this may be a coincidence and we do not claim any physical significance, although we also have no reason to reject any. We see that the scaling of $k_{coll}$ with $\talpha$ depends strongly on which inertial range one is looking at, a reflection of the difference of behavior in the two inertial ranges. Our results tell us that the scales larger than the forcing couple more slowly than those at scales smaller than the forcing.


\bibliographystyle{jfm}
\bibliography{2DPIMHD_refs}

\end{document}